\pdfoutput=1

\documentclass[sigconf]{acmart}

\usepackage{balance}

\usepackage{xspace}
\usepackage{subcaption}
\usepackage{multirow}
\usepackage{enumitem}
\usepackage{listings}
\usepackage{makecell}
\usepackage{colortbl}

\newcommand{\RaggedRight}{%
  \rightskip=0pt plus 1fil\relax 
  \hyphenpenalty=50 
  \exhyphenpenalty=50 
  \tolerance=1000 
  \emergencystretch=3em 
  \raggedright 
  \hbadness=1000 
}

\makeatletter
\AtBeginDocument{%
  \begingroup
  \small
  \let\tmp@n@s\f@size
  \let\tmp@n@b\f@baselineskip
  \normalsize
  \let\tmp@s@s\f@size
  \let\tmp@s@b\f@baselineskip
  \xdef\semismall@size{\fpeval{(\tmp@n@s+\tmp@s@s)/2}}%
  \xdef\semismall@baselineskip{\fpeval{(\tmp@n@b+\tmp@s@b)/2}}%
  \endgroup
}
\newcommand{\semismall}{\fontsize{\semismall@size}{\semismall@baselineskip}\selectfont}

\newcommand*\circled[1]{\raisebox{.5pt}{\large\textcircled{\raisebox{-.5pt} {\semismall\textsf{#1}}}}}

\newcommand{\wrap}[1]{{\ttfamily{#1}}}

\newif\ifdisabletrackchanges
\disabletrackchangesfalse 
\disabletrackchangestrue 

\definecolor{darkpastelred}{rgb}{0.92, 0.2, 0.18}
\definecolor{burntorange}{rgb}{0.75, 0.38, 0.05}

\ifdisabletrackchanges
  \newcommand\del[1]{%
  }%
  \newcommand\add[1]{#1}
  \colorlet{AddColor}{black}
  \newcommand\addImg[1]{#1}
  \newcommand\addDiscuss[1]{}
  \newcommand\deleteDiscuss[1]{}
\else
  \newcommand\del[1]{%
    \textcolor{darkpastelred}{\sout{#1}}%
  }
  \newcommand\add[1]{%
  \textcolor{Cerulean}{#1}%
  }
  \colorlet{AddColor}{Cerulean}
  \newcommand\addImg[1]{\setlength{\fboxrule}{1pt}\fcolorbox{Cerulean}{white}{#1}}
  \newcommand\addDiscuss[1]{\textcolor{burntorange}{\uuline{#1}}}
  \newcommand\deleteDiscuss[1]{\textcolor{burntorange}{\uuline{#1}}}
\fi

\newcommand\edit[2]{\del{%
#1%
}\add{#2}}

\newcommand{\toolName}{Screen\-Au\-dit\xspace}

\AtBeginDocument{%
  }

\copyrightyear{2025}
\acmYear{2025}
\setcopyright{cc}
\setcctype{by}
\acmConference[CHI '25]{CHI Conference on Human Factors in Computing Systems}{April 26-May 1, 2025}{Yokohama, Japan}
\acmBooktitle{CHI Conference on Human Factors in Computing Systems (CHI '25), April 26-May 1, 2025, Yokohama, Japan}\acmDOI{10.1145/3706598.3713797}
\acmISBN{979-8-4007-1394-1/25/04}

\begin{document}

\title{\toolName: Detecting Screen Reader Accessibility Errors in Mobile Apps Using Large Language Models}

\author{Mingyuan Zhong}
\orcid{0000-0003-3184-759X}
\affiliation{%
  \department{Computer Science \& Engineering}
  \institution{University of Washington}
  \city{Seattle}
  \state{WA}
  \country{USA}
}
\email{myzhong@cs.washington.edu}

\author{Ruolin Chen}
\orcid{0009-0006-7981-1966}
\affiliation{%
  \department{The Information School}
  \institution{University of Washington}
  \city{Seattle}
  \state{WA}
  \country{USA}
}
\email{ruolin@uw.edu}

\author{Xia Chen}
\affiliation{%
  \institution{Carnegie Mellon University}
  \city{Pittsburgh}
  \state{PA}
  \country{USA}
}
\email{xiac@andrew.cmu.edu}

\author{James Fogarty}
\orcid{0000-0001-9194-934X}
\affiliation{%
  \department{Computer Science \& Engineering}
  \institution{University of Washington}
  \city{Seattle}
  \state{WA}
  \country{USA}
}
\email{jfogarty@cs.washington.edu}

\author{Jacob O. Wobbrock}
\orcid{0000-0003-3675-5491}
\affiliation{%
  \department{The Information School}
  \institution{University of Washington}
  \city{Seattle}
  \state{WA}
  \country{USA}
}
\email{wobbrock@uw.edu}

\renewcommand{\shortauthors}{Zhong et al.}

\begin{abstract}
Many mobile apps are inaccessible, thereby excluding people from their potential benefits. Existing rule-based accessibility checkers aim to mitigate these failures by identifying errors early during development but are constrained in the types of errors they can detect.
We present \toolName, an LLM-powered system designed to traverse mobile app screens, extract metadata and transcripts, and identify screen reader accessibility errors overlooked by existing checkers. We recruited six accessibility experts including one screen reader user to evaluate \toolName's reports across 14 unique app screens. Our findings indicate that \toolName achieves an average coverage of \edit{69.9\%}{69.2\%, compared to only 31.3\% with a widely-used accessibility checker}.
\del{Feedback from the experts suggests that \toolName's features could benefit app developers in real-world settings.}
Expert\edit{s expressed}{\ feedback indicated} that \toolName delivered higher-quality feedback and addressed more aspects of screen reader accessibility compared to existing checkers\add{, and that \toolName would benefit app developers in real-world settings}.
\end{abstract}

\ccsdesc[500]{Human-centered computing~Accessibility systems and tools}

\keywords{Mobile accessibility, large language models, accessibility audit.}

%

\maketitle

\section{Introduction}
    Mobile applications frequently fail to meet accessibility standards \cite{ross_examining_assets18,yan_current_state_accessibility,ross_epidemiology_taccess20,fok_large-scale_chi22}, rendering app contents and functionality inaccessible to people with disabilities.
Despite efforts to promote developer awareness~\cite{apple_guidelines,material_design_guidelines,w3c_guidelines} and availability of developer toolkits~\cite{a11y_inspector, a11y_scanner, axe}, 
a recent large-scale longitudinal study~\cite{fok_large-scale_chi22} found no significant improvements in mobile app accessibility.
Meanwhile, an industry survey~\cite{levelaccess2024report} found that 72\% of developers are \textit{not} certain that their company's digital services are accessible.

Contributing to the lack of accessibility improvement and developer confidence may be the limited accessibility error coverage in current automated checkers (e.g., Google's Accessibility Scanner)~\cite{Carvalho_2018_a11yproblems, mateus2020eval, a11y_scanner}.
For instance, Carvalho et al. found that blind and partially sighted users encountered 40 distinct types of accessibility problems across eight mobile apps and websites~\cite{Carvalho_2018_a11yproblems}, yet  
automated checkers detected only four of these problems~\cite{mateus2020eval}.
Our own analysis of student developer usage of standard Android development tools revealed that automated checkers were perceived as ineffective and inconsistent and were challenging to use and to interpret.

The importance of accessibility-first design has long been recognized for creating inclusive user interfaces~\cite{stephanidis1998universal, accessibility_first}.
This approach aligns with guidelines suggesting that prioritizing accessibility from the initial stages of development is crucial~\cite{apple_guidelines,google_guidelines,pellegrini2020prioritize,shirogane2011method}.
Although automated checkers are widely adopted in the industry~\cite{levelaccess2022report} and are valuable for identifying issues early and cost-effectively, their current limitations call for improvements in coverage and quality.
Recent advancements in large language models (LLMs)~\cite{attentionisallyouneed,openai_gpt-4_2023} demonstrate unprecedented UI understanding capabilities~\cite{huang2024promptrpa, zhang2023appagent, wang2024mobile} and have been adopted to automatically operate mobile apps, surfacing interface and accessibility issues for software quality assurance~\cite{liu2023chatting,taeb24axnav}.

\begin{figure*}[ht]
  \includegraphics[width=0.75\textwidth]{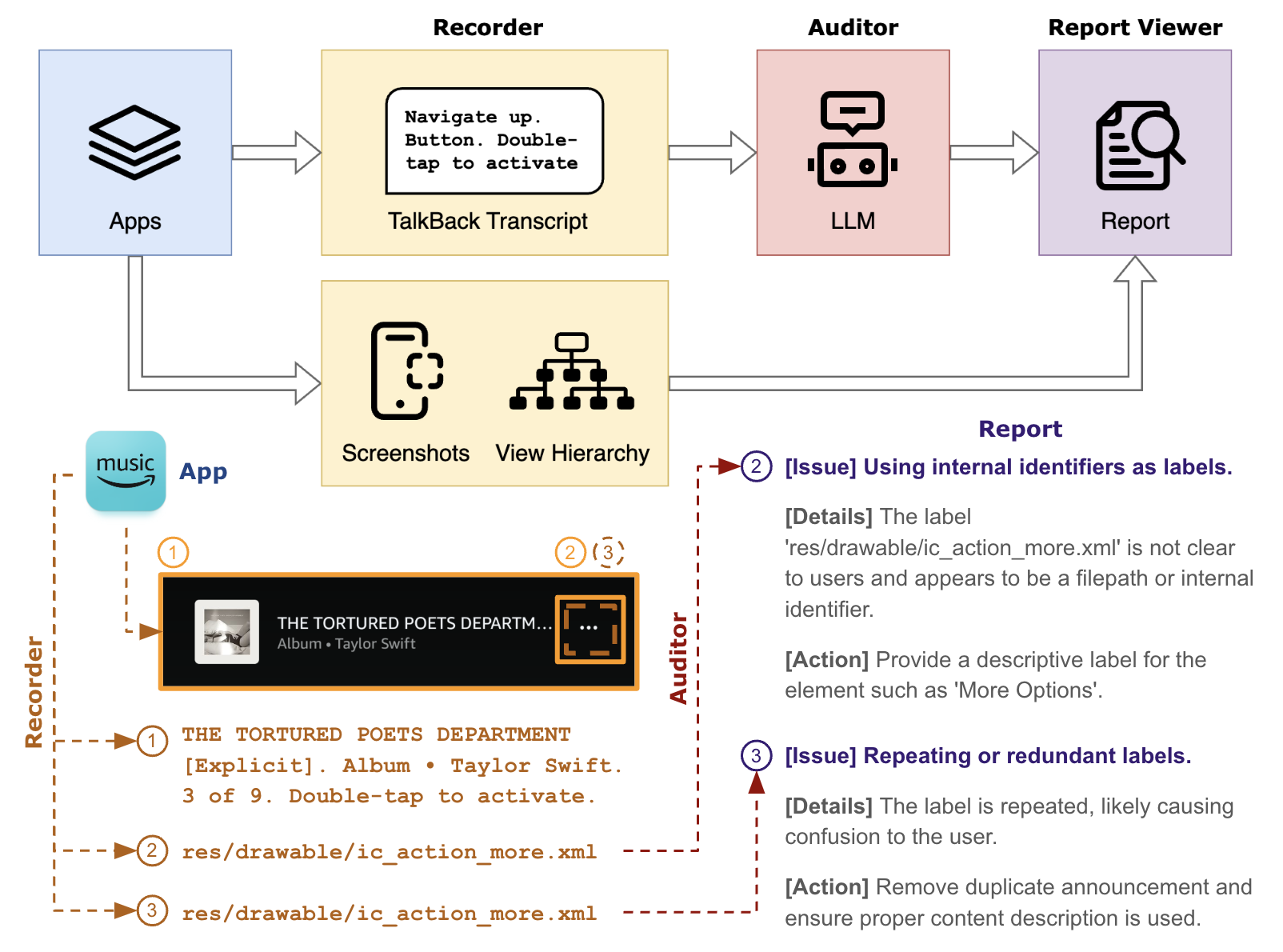}
  \caption{\toolName captures accessibility metadata from an app screen, including TalkBack transcripts. An LLM is used to evaluate potential screen reader accessibility errors, which are presented in a report. In the example from Amazon Music, \toolName identifies the errors of using internal identifiers and redundant labels and provides actionable advice.}
  \label{fig:system}
\end{figure*}

Inspired by these results, we present \textit{\toolName}, an exploratory accessibility checker for Android apps targeted at identifying screen reader accessibility errors and surfacing them to app developers.
\toolName aims to enhance the quality, coverage, and interpretability of current automated accessibility checkers by incorporating large language models (LLMs).
In contrast to existing accessibility checkers, \toolName automates TalkBack (Android's default screen reader) to traverse and collect speech output from app screens, and then utilizes OpenAI's GPT-4o~\cite{openai_gpt-4_2023} to interpret these outputs and identify common accessibility errors (see Figure~\ref{fig:system}).
We prompt the LLM with general instructions about TalkBack output and accessibility failures, without limiting to particular guidelines. Shaped by developer feedback, our reports include screen reader outputs, identified issues and their explanations, and suggested actions for repair.

We evaluated \toolName reports through an expert study. Six accessibility experts (one blind screen reader user and five sighted experts) analyzed a total of 14 mobile app screens independently without prior exposure to our reports.
Findings show a strong correlation between expert assessments and the reports generated by \toolName.
Of 163 errors identified by experts, \toolName uncovered 69.2\%, with a precision of 71.3\%.
Notably, \toolName identified at least four additional issues initially overlooked by the experts but confirmed upon review.
Although limitations exist, experts found \toolName to be effective and easy to use.
We constructed an accessibility error dataset based on expert audits and evaluated performance of different prompting strategies.
A prompt that provided general accessibility guidance and encouraged LLM's contextual lookup yielded the best result.

Our goal in creating \toolName is not to replace user testing during development but to complement it.
Early and frequent testing is essential.
\toolName aims to provide immediate, actionable feedback for developers, allowing user testing sessions to focus on more complex, nuanced issues that are best understood through user interaction.

In summary, our main contributions are:
\begin{enumerate}[topsep=0pt]
    \item The \toolName system, an accessibility audit tool for developers that captures screen reader announcements and automatically generates accessibility reports with explanations using an LLM.
    \item A study with accessibility experts that demonstrates the coverage and usefulness of \toolName.
    \item An evaluation of different LLM prompts and their effects on the detection of accessibility errors.
\end{enumerate}

\section{Related Work}
    Our research builds on related work in automated accessibility evaluation tools 
and their limitations
and related work in automated assessments using large language models (LLMs).

\subsection{Automated Accessibility Evaluation \add{Tools}}

Automated evaluation tools \edit{has}{have} been extensively used \edit{to automate the process of}{to} identify\del{ing} accessibility issues \add{in software~\cite{levelaccess2022report}}. Compared to manual inspection, \edit{it has}{they offer} advantages in efficiency and scalability. Automated accessibility \edit{testing}{evaluation} tools can be broadly categorized into two types: code-level linters and run\del{-}time analyzers. 

Code-level linters\add{, such as for Android Studio}~\cite{accesslinter}\add{, React Native~\cite{reactnative_eslint}, GitHub~\cite{accesslint}, and Deque's Axe Accessibility Linter~\cite{deque2024axelint},  provide basic accessibility checking directly in common developer tools and platforms.} Linters \edit{are used during app development,}{almost always} employ \add{rule-based} static analyses to check for potential accessibility errors and \edit{suggesting}{highlight} problems in\del{ the} code \add{during app development.
However, they can only check a limited set of guidelines~\cite{accesslinter, axelinterrules, reactnative_eslint, santacruz2020flutter} and are unable to examine dynamically generated contents or components during runtime.}

Runtime \edit{tools}{accessibility checkers complement linters in the developer workflow by detecting problems in running applications.}\edit{, like}{\ One type of accessibility checker analyzes accessibility errors on individual app screens or pages, such as} Google’s Accessibility Scanner~\cite{a11y_scanner}, Apple’s Accessibility Inspector~\cite{a11y_inspector}, \add{Deque's} Axe~\cite{axe}, and WAVE~\cite{wave}. 
They inspect and analyze every element \edit{and}{in the UI} hierarchy of each screen, offering \add{additional} insights beyond static analysis\add{~\cite{silva2018survey}}.
\add{However, these checkers still require manual effort in navigating to and interacting with a specific screen under test.}

\add{To address this, another line of work combines the analysis of individual screens with programmatic exploration of apps (i.e.,~crawling).}
\add{MATE~\cite{eler2018mate} randomly activates interactive elements in an app to visit different app states and check for accessibility errors heuristically.}
\add{To simulate the experience of assistive technology users, Groundhog~\cite{Salehnamadi2023groundhog} invokes assistive services during automated exploration to test whether UI elements can be focused and activated.
Similarly, BAGEL~\cite{chiou2023bagel} automatically performs keyboard navigation on a webpage to detect accessibility issues in navigation.}
\edit{Such tools can also be combined with automated crawlers to automatically cover many screens within an app}{Some of these automated exploration techniques have been adopted in large-scale assessments of mobile app accessibility}~\cite{fok_large-scale_chi22, yan_current_state_accessibility, chen2022accessible}.

\add{Recent research has started to study and mitigate the problems encountered during app crawling.
For example, apps often have designs that hinder programmatic access to certain screens, such as those behind a login screen, an ad, or pop-up windows~\cite{fok_large-scale_chi22}.
To visit potentially blocked screens, juGULAR~\cite{amalfitano2019combining} uses a statistical classifier to detect such states and replays recorded event sequences to bypass them.
Another problem arises when presenting accessibility issues to developers: issues identified across an app can be redundant and overwhelming, as screens are often re-visited and UI components re-used.
To de-duplicate and summarize issues, Swearngin et al.~\cite{swearngin2024towards} developed a screen grouping model that supports the reporting of unique accessibility issues in apps.}

\add{Still, limitations exist in all current automated accessibility evaluation tools,
as a survey found existing tools covered only 8 of 64 accessibility guidelines from BBC and WCAG 2.0~\cite{silva2018survey}.
Deque's own assessment claimed that their tool covered 16 of the 50 Success Criteria under WCAG 2.1 Level AA~\cite{deque2024coverage}.} 
In an evaluation of blind and partially sighted people's usage of mobile websites and apps, users encountered 40 distinct types of accessibility problems~\cite{Carvalho_2018_a11yproblems}, yet current automated checkers detected only four of these problems~\cite{mateus2020eval}.
Many frequent problems \add{blind and low vision (}BLV\add{)} users encounter, such as inappropriate feedback, unclear or confusing functionality, inconsistent navigation, are not covered by existing checkers~\cite{Carvalho_2018_a11yproblems}.

\add{The scope of our exploratory work here focuses on individual app screens, consistent with tools widely adopted in the industry~\cite{a11y_scanner, a11y_inspector, axe}.
We perform automated crawls within the scope of an app screen, similar to prior work~\cite{Salehnamadi2023groundhog, chiou2023bagel}, to capture the perspective of assistive technology users.}
We expand on existing run\del{-}time accessibility evaluation tools by extending the types of accessibility errors that they can detect.

\subsection{LLM-Supported Automated Assessments and UI Understanding}
Large language models (LLMs) have demonstrated proficiency in conducting automated assessments that traditionally rely on human intervention. Zhang et al.~\cite{zhangRepairingBugsPython2022} developed an LLM-based system for grading programming assignments that identified syntactic and semantic errors and provided feedback comparable to human instructors.
Liang et al.~\cite{liangCanLargeLanguage2023a} showed LLMs were also able to produce feedback on academic papers that resembled that of human peer reviewers, with over 50\% of participants perceiving AI-generated feedback to be more valuable than  feedback from at least some human reviewers.
Closer to our current research, Duan et\del{.} al\add{.}~\cite{duanUICritEnhancingAutomated2024} demonstrated LLMs could evaluate and generate feedback for UI designs, and their performance can be significantly improved via few-shot and visual prompting techniques.
\add{Wu et al. trained UIClip~\cite{wu2024uiclip}, a vision-language model that quantifies UI design quality that can improve LLM UI code generation.}

LLMs also show promising results in other UI understanding tasks, such as generating help~\cite{zhong_helpviz_chi21}, extracting macros~\cite{huangAutomaticMacroMining2024}, and autonomously executing tasks as agents~\cite{huang2024promptrpa,zhang2023appagent,wang2024mobile,wen2024autodroid}.
Such UI understanding and decision\add{-making} capabilities have been utilized in performing automated GUI and accessibility tests.
GPTDroid~\cite{liu2023chatting} uses an LLM to explore mobile apps and automatically generate GUI tests.
AXNav~\cite{taeb24axnav} adopts a similar approach by using LLMs to perform accessibility tests based on natural language instructions and heuristically flag accessibility issues.

These efforts demonstrate strong capability of LLMs in understanding and \edit{operating}{interacting with} mobile UIs.
However, such UI knowledge remains largely untapped for analyzing accessibility issues themselves.
\toolName aims to leverage \edit{such}{these} UI understanding and assessment capabilities to identify and explain accessibility problems in mobile apps.

\section{Formative Analysis on Student Developer Perspectives} \label{sec:dev-survey}
    \addDiscuss{[Revision Note] We added study details to this section.}

Although existing research has established the inadequacy of existing accessibility checkers, \add{not enough is known about} the perspectives of the users of these developer tools (i.e.,~the developers themselves).
To this end, we explored the responses to \add{an assignment's} reflection questions from student developers in an undergraduate course on Android programming, 
\add{designated as an intermediate-level ``core course'' in our university's computer science major. In the assignment,} students were tasked with evaluating and repairing accessibility errors using the Android Accessibility Scanner~\cite{a11y_scanner} and TalkBack~\cite{talkback} \add{for apps that they created}.
\add{Full text of this assignment and the reflection questions is available in Appendix~\ref{sec:appendix-assignment}}.

\add{The questions analyzed were assigned as part of the regular curriculum of the course.
Prior to this assignment, students had received seven weeks of course materials on Android development, including two lectures on accessibility. They also had developed six prototype Android apps covering different aspects of programming interactive applications.
Our formative analysis of student feedback is based on responses stripped of all personally identifiable information (PII). We received a determination that this study was not human subjects research from our University's Institutional Review Board (IRB) because of the exclusion of all PII.}

We chose to analyze student \edit{responses because their reflections}{reflections because they} mirrored the initial experiences and challenges that many developers have when checking and repairing accessibility failures.
The structured academic setting ensured that students received uniform instructions and operated under similar conditions.
Student-identified accessibility failures
covered 11 of the 15 most frequently encountered problem types by \edit{blind screen reader users}{blind people who use screen readers} in prior work~\cite{Carvalho_2018_a11yproblems, mateus2020eval}.
Detailed procedure and analysis results are available in Appendix~\ref{sec:appendix-qualitative}.

\subsection{Feedback for Current Developer Tools}
Using standard thematic analysis~\cite{braun2006thematic}, we analyzed student reflections on the problems of current developer tools. We also sought to understand the features they desired to better support accessible app development.

\subsubsection{Problems}
Our analysis revealed three themes in student problems with current developer tools:
P1:~\textit{Limited detection scope and inconsistency in Accessibility Scanner}, where students noticed unreported problems, and identified inconsistent or false positive issues.
P2:~\textit{Difficulty operating TalkBack}, where students reported encountering bugs and difficulty setting up TalkBack and navigating using gestures and hearing announcements.
P3:~\textit{Lack of developer support in all developer tools}, where students had difficulty with locating relevant tools and interpreting their results \add{on potential accessibility issues}.

\subsubsection{Features}
Our analysis further revealed three main features students wanted in developer tools to enhance accessible app development:
F1:~\textit{Integrate accessibility tool output and provide accessibility alerts during development}: Accessibility Scanner results (alerts and warnings) and TalkBack announcements should be integrated with Android Studio.
F2:~\textit{\edit{Provide better explanations and developer education of issues}{Enhance explanations of issues and improve developer education}}: Tools should provide examples and explanations for potential accessibility issues, along with tutorials and learning materials.
F3:~\textit{Conduct code-level scanning and provide suggestions}: Tools should offer intelligent code analysis to flag potential accessibility issues and and provide refactoring suggestions.

\subsection{Design Implications}
Our findings show that developers, especially those new to accessibility, encounter significant challenges that current development tools fail to address.
We identify five design implications (D1--D5) for an accessibility checker based on the above findings and problems identified by prior work.
As an exploration of incorporating LLMs to support accessibility checking, \toolName focuses on the identification and explanation of errors, and implements three of the design implications (D1--D3).
D4\edit{,}{\ and} D5 \edit{warrant}{require} additional research and development \edit{and provide opportunity for}{in} future work~\add{ on developer tools}.

\subsubsection{D1: Expand the Coverage of Current Accessibility Checkers}
Developers and prior work~\cite{Carvalho_2018_a11yproblems} identified the limited coverage of error types supported by existing checkers, limiting their benefits and effectiveness (P1). We design \toolName to expand coverage while recognizing the benefits of rule-based checking for its efficiency, reproducibility, and usefulness for certain error types (e.g., missing labels, color contrast, target size). Therefore, \edit{our current system}{\toolName} is intended to operate alongside existing checkers and tools, while future iterations can directly incorporate rule-based checking features~\cite{google-atf}.

\subsubsection{D2: Integrate Screen Reader Output}
Developers who had no prior exposure to screen readers have difficulty in setting up and operating TalkBack even after tutorials given in lectures (P2).
Developers also \edit{wished}{hoped} to see TalkBack's output integrated into developer tools \edit{so that they can quickly identify or confirm problems}{in support of identifying or confirming problems} without context-switching (F1). \edit{We}{\toolName} automatically capture\add{s} TalkBack outputs for the target screen and include\add{s} them in the generated accessibility report for quick reference.

\subsubsection{D3: Provide Explanations and Actions for Errors Detected}
Developers had difficulty interpreting the Accessibility Scanner's outputs as certain terms were considered overly technical (P3); developers desired better explanations with examples \add{of accessible designs} (F2).
We designed \toolName to generate concise explanations with suggested actions for each error \edit{identified}{detected}.
We anticipate that future iterations of our tool will provide direct links to relevant tutorials or guidelines to support developer education.

\subsubsection{Additional Design Implications} \label{sec:design-additional}
We further identify two design implications: 
D4: \textit{Integrate accessibility checker features directly into Android Studio or other IDEs} (P3, F1),
and D5: \textit{Provide code-level analysis and suggestions for accessibility} (P3, F3).
We did not implement these features but acknowledge their promise as potential future work, whether by us or by others.

\section{\toolName: System Design}
    \addDiscuss{[Revision Note] We rewrote Section~\ref{sec:report-viewer} to clarify the functionality of Report Viewer, and provide concrete examples for interpreting the results. We added an example to Figure 1 and moved it to the beginning of the paper (in Section 1) to better illustrate how \toolName works.}

We created \toolName based on the design implications identified in our formative analysis and in prior work.
\toolName is intended to \edit{be used}{function} similar to current accessibility checkers: a developer opens an app screen to test on their phone or emulator and initiates a scan with \toolName.
After the scan is complete, the developer can view the report through a web portal on their computer.

The \toolName system consists of three main components: 
a~\textit{Recorder} that traverses a mobile app screen and captures necessary metadata including its TalkBack output, 
an \textit{Auditor} that evaluates potential screen reader accessibility errors on the screen using an LLM, 
and a \textit{Report Viewer} that displays a report highlighting detected errors and provides explanations.
Figure~\ref{fig:system} illustrates \toolName's system pipeline.

\subsection{Recorder}
\toolName collects metadata \add{of the screen being scanned} with its Recorder.
\edit{In particular, we are interested in}{One important goal of the Recorder is to} collect TalkBack's spoken feedback so the Auditor can analyze the same announcements a \edit{screen reader user would hear}{user would hear on screen readers}.
\add{This also fulfills design implication D2. In contrast,}
traditional rule-based checkers examine a static snapshot of a screen, which ignores \add{its} dynamic aspects (e.g., focus order, scrollable views) and cannot effectively use the context provided by neighboring elements.

\subsubsection{Using Gestures to Control TalkBack and Traverse a Screen}
Our initial attempt to capture TalkBack output was based on the TalkBack logs\footnote{These logs can be accessed by setting TalkBack's log level to verbose and searching for \textit{SpeechControllerImpl}'s speaking entries.} and sending SWIPE\_RIGHT gestures to the device. This approach captured accurate \edit{speech outputs}{announcements} but had three main drawbacks:

\begin{enumerate}
    \item \textit{The time needed to traverse a screen is long}.
        In our testing, even with text-to-speech (TTS) set at max speed, the average time for capturing one element was three seconds.
        We observe between 10 and 40 elements in most app screens before any scrolling occurs.
        Therefore, we \add{typically} need between 30 and 120 seconds to capture a screen's TalkBack output, which reduces the efficiency of the entire auditing process.
        \edit{The reason for the delay}{Three factors contribute to the overall delay: 
        \add{(1)} The capture duration depends on \edit{the length of content}{content length, which can range from under one second to tens of seconds}.
        (2) A 0.5-second delay} is \add{needed} to ensure that we capture an element's usage hint, which has a built-in delay of 0.4 seconds \add{from TalkBack}.
        \add{A usage hint is necessary for conveying an element's accessibility role, such as ``Double-tap to activate.''}
        \add{(3)} Furthermore, an extra $\mathord{\sim}0.4$ seconds is needed to send a swipe gesture \add{to the device} and wait for \edit{its}{a} response ~\add{from TalkBack}.

    \item \textit{Lack of standardization of captured results}.
        As each developer can choose to customize their TalkBack settings, there is no guarantee \del{of consistency }when deploying \del{the }\toolName \del{system }that \edit{their}{a developer's} settings will be the same as ours.
        For example, usage hints can be hidden \edit{if corresponding setting}{if turned off} in TalkBack\add{'s settings}\del{ is turned off}.
        This inconsistency will \add{impact the audit accuracy by} misleading the Auditor to believe that \add{all} usage hints are missing.

    \item \textit{Focus indicator is captured}.
        TalkBack shows a green focus indicator \edit{to indicate which}{for the} element \edit{has}{with} screen reader focus.
        Although this feature is valuable for \add{users with} low vision \del{users }and for testing \add{screen user support}, the indicator can distract \edit{our user}{developers if captured as part of the screenshot in \toolName's report}.
\end{enumerate}

To address \edit{the above}{these} issues, we created a custom fork of TalkBack based on its open source implementation~\cite{google_talkback_source_2024} that preserves its normal speech output behaviors, but with additional capabilities:
(1) Efficient speech capture and traversal using TalkBack's internal pipeline.
\add{Specifically, whenever the text for an announcement is created, our implementation skips TTS and directly proceeds to the next focusable element.}
(2) Settings are stored \edit{independent from}{independently of} TalkBack.
(3) The focus indicator can be hidden.
Our improvements reduced the capture time per element to around 0.15 seconds (\add{limited by} the internal processing \edit{delay from}{latency of} TalkBack) while maintaining the ability to capture usage hints, which means \add{typically} a screen takes between 1.5 and 6.0 seconds to capture, regardless of \add{spoken} content length.

\subsubsection{Capturing TalkBack Transcripts and Screen Snapshots}~\label{sec:capture}
\edit{We}{The Recorder} capture\add{s} TalkBack \edit{speech outputs}{announcements} by sequentially focusing on each screen reader focusable element, starting from the default element of initial focus.
Similar to prior work~\cite{taeb24axnav} and in keeping with the Accessibility Scanner, we limit the total number of elements traversed to the first screen or 40 elements, to prevent infinite loops or scrolls.
In addition, \edit{we}{the Recorder} capture\add{s} the corresponding screen snapshot with screenshot and view hierarchy data\add{,} similar to Fok et al.~\cite{fok_large-scale_chi22}.
\edit{We associate the}{Each} TalkBack announcement \add{is associated} with its corresponding element on screen by comparing their bounding boxes.

While traversing a screen, \add{dynamic UI} changes in content or layout may occur, \add{such as when scrolling or loading new contents}.
The Recorder monitors {\small\wrap{TYPE\_VIEW\_SCROLLED}} and {\small\wrap{TYPE\_WINDOW\_CONTENT\_CHANGED}} events from the Android system's Accessibility Service to determine if a potential change has occurred and capture a new snapshot.
Redundant snapshots \add{are removed} if there \edit{are}{is} no pixel-level difference.

\subsection{Auditor}
We focus the design and implementation of the Auditor on fulfilling design implications D1 and D3, \edit{to expand}{expanding} the coverage of existing accessibility checkers and \edit{to provide}{providing} explanations and actions for errors detected.
To implement these, we \edit{provide}{use} the TalkBack transcript as part of our custom prompt to request feedback from the LLM, and prompt the LLM to provide explanations and actions for \edit{errors}{each error}.
The Auditor sends the prompts to GPT-4\add{o~\cite{gpt4o} (a derived model of GPT-4~\cite{openai_gpt-4_2023})} for completion.

\subsubsection{Initial Observation}
Our initial attempt identified \edit{main}{the} strength and weakness of GPT-4\add{o}'s assessments\edit{, when directed to evaluate the}{\ evaluating TalkBack} transcripts for screen reader accessibility without \edit{further}{additional} instructions.
The main strength\edit{s are}{\ is} its ability to infer information based on surrounding context from the transcript.
For example, in a food delivery app there are two consecutive \edit{speech outputs}{announcements}: ``30 min'' and ``Delivery time.''
The LLM is able to associate them together and provide the following suggestion: ``Consider grouping `30 min' and `Delivery time' together to avoid redundancy and to provide clear context.''

\edit{Its}{The} main weakness\edit{es are}{\ is} \edit{its}{the model's} lack of specific knowledge of \add{TalkBack and} mobile accessibility.
    For example, the output ``Selected, Home. Tab 1 of 4. Double-tap to activate'' follows the default announcement of TalkBack in the order of \textit{state}, \textit{label}, \textit{role}, \textit{usage hint}.
    However, such formatting can result in false positives from the LLM, where the model move\add{s the} \textit{state} \edit{back}{(``Selected'') to the end}, especially when the label and usage hints are more complex.
    In another example, the LLM claims this output ``\$0 Delivery Fee on \$15+'' is not clear because it did not indicate if it is plain text or an actionable item. \add{However this assessment} \edit{contrary to accessibility guidelines (}{contradicts} WCAG 4.1.2\del{)}, which do\add{es} not require textual labels to be indicated for conciseness.

\subsubsection{Prompt Structure} \label{sec:prompt}

\begin{table*}[ht]
\caption{Prompt Structure for Analyzing TalkBack Transcripts}
\label{tab:prompt}
\renewcommand{\arraystretch}{1.35}
\centering
    \begin{tabular}{c c p{11cm}}
        \toprule
        \textbf{Section} & \textbf{Context} & \textbf{Prompt} \\
        \midrule
        \rowcolor{gray!10} \textit{Introduction} & Static & You are examining the accessibility of an app based on transcripts of TalkBack. You will see a transcript of what a screen reader user will hear when they use an app. Please analyze if each interaction reflected in this transcript is accessible. ... \\

        \textit{Accessibility} & Static & \# Accessibility Basics \newline
        \verb| |\ \ \ We want descriptions to be informative and concise. Consider if the transcript for each element conveys its content or purpose appropriately. Most important information should appear first. ... \\

        \rowcolor{gray!10} \textit{Instruction} & Static & \# Your Task \newline
        \verb| |\ \ \ Follow these steps to evaluate the accessibility of the transcript:\newline
        \verb| |\ \ \ Step 1 - Look at the big picture. Consider each transcript entry in relation to the elements before and after it. Is this element related to nearby elements? ...\\

        \textit{Transcript} &  Dynamic & app: "...", \newline transcripts: [ ... \newline
            \verb| |\ \ \ \{ index: 6, transcript: "\textit{Image Search. Button. Double-tap to activate}"\}, \newline
            \verb| |\ \ \ \{ index: 7, transcript: "\textit{Appliances}"\},  
        ... ] \\
        \bottomrule
    \end{tabular}
\end{table*}

To improve the accuracy of the model's accessibility assessments, we experimented with different ways of including accessibility guidelines and principles.

Table~\ref{tab:prompt} details the prompt structure we adopted in our analysis.
Our prompt consists of four sections: \textit{Introduction}, \textit{Accessibility} overview, \textit{Instruction}, and a TalkBack \textit{Transcript} for the screen being evaluated.
Except for the Transcript section that includes dynamic information of the transcript being analyzed, all other sections are static.
We design our prompt based on our initial observations and commonly encountered accessibility problems discussed in prior work.
Our full prompt can be found in Appendix~\ref{sec:appendix-prompt}.

The Introduction section briefly describes the task.
In the Accessibility overview section, we provide fundamental accessibility guidelines to TalkBack in one of two ways: (1) a \textit{General} guideline independently written by two of the researchers and then compiled together, and 
(2) a \textit{WCAG} 2.1-based list of accessibility guidelines, with items that are not relevant to transcript analysis removed.
We compare these against a \textit{Base} condition\edit{ where the}{, which lacks the} Accessibility overview section\del{ is removed}.

In the Instruction section, we provide chain-of-thought instructions~\cite{wei2023chainofthought} of accessibility evaluation requirements\edit{.
We would like to understand if specifically asking the model to check for context will improve its ability to do so.}{, with the aim of understanding whether the model can better check for context when explicitly instructed to do so.}
We include two prompts to evaluate this:
(1) a \textit{Basic} instruction that asks the model to evaluate the transcript for its accessibility, 
and (2) a \textit{Contextual} instruction that has an additional step that asks the model to look at nearby elements for context.

Finally, the \del{static prompt is appended the dynamic }Transcript section \add{is appended to the end} to form a complete prompt.
As these designs were made through informal testing, we evaluate all prompt designs after collecting \add{a} ``ground truth'' accessibility error \edit{labels after}{dataset from} experts (see Section~\ref{sec:prompt-eval}).

\subsection{Report Viewer} \label{sec:report-viewer}

\begin{figure*}[ht]
    \centering
    \fcolorbox{gray}{white}{\includegraphics[width=0.8\textwidth]{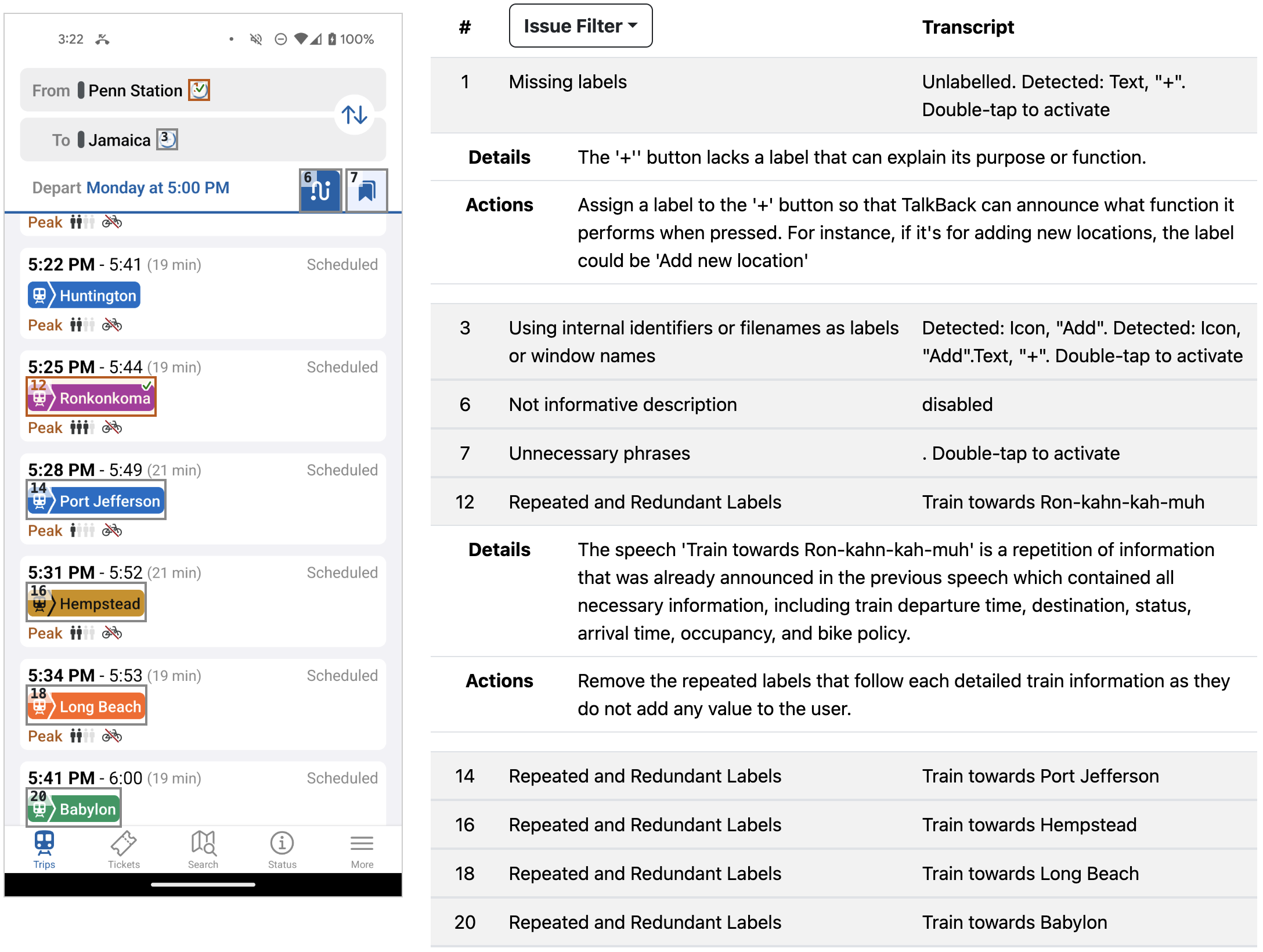}}
    \caption{The Report Viewer of \toolName displaying \del{the }accessibility errors detected in a mobile application. The report entries for item\del{s} 1 (unlabeled add button for ``From'' station) and \add{item} 12 (a redundant label) are expanded with their details shown. \add{The ``Issue Filter'' menu (top) allows filtering of errors by type.}}
    \label{fig:reportviewer}
\end{figure*}

The Report Viewer consists of screenshots captured by the Recorder and evaluations generated by the Auditor in an expandable list view (see Figure~\ref{fig:reportviewer}). 
On the left, a screenshot highlights detected accessibility errors. Locations of the elements are determined based on the view hierarchy data captured by the Recorder. 
Hovering (or pressing the tab key) over each element will show its TalkBack transcript and detected error. 
On the right, a list of detected accessibility errors are shown with brief descriptions at a glance. 
An element can be selected on the screenshot or in the list.
According to the element selected, if an error exists, the list view expands to display its TalkBack transcript with details and recommended actions on error handling for the developers.

\add{The list view in Figure~\ref{fig:reportviewer} shows details on two exemplar accessibility errors detected by \toolName:}
\begin{enumerate}
    \item \add{The error at index 1 shows the ``add'' icon does not have an appropriate label and \toolName suggests to add a label ``Add new location''.
    The transcript shows that the button is entirely unlabeled by the developer, and that TalkBack attempts to mitigate the problem by auto-generating a label~``+''.
    Therefore, the button here violates WCAG~1.1.1, which requires non-text contents to have text alternatives.
    The label generated by TalkBack violates WCAG~2.4.6, which requires labels to be descriptive.}

    \item \add{The error at index 12 shows \toolName identifies redundant announcement for the train element.
    Immediately prior to this, the element at index 11 reads ``{\small\texttt{5:25 PM peak train towards Ron-kahn-kah-muh. Scheduled}} ...'' \allowbreak(not shown in the screenshot).
    \toolName identifies the train information to have existed in the previous speech, and suggests removing the label.
    This type of error has been identified in prior work~\cite{Mehralian2023overly} as ``overly perceivable.''
    Specifically, this label violates WCAG~1.3.1, which requires information to be equally perceivable to all users.
    }
\end{enumerate}

\section{\add{Accessibility Error Dataset Construction and} Expert Evaluation}

We recruited accessibility experts with significant experience with screen readers to examine the report produced by \toolName.
We have two main goals in conducting this study: (1)~to construct a ground-truth dataset of accessibility errors to evaluate the coverage and accuracy (recall and precision) of our system, and (2)~to obtain expert subjective opinions on \toolName's reports.
\add{We received approval from our University's Institutional Review Board (IRB) prior to the study.}

\add{In this section, we first introduce how we selected the Android apps for the study, then the expert profiles and study procedure.
After that, we provide details on our analysis and study results.
Finally, we illustrate \toolName's success and failure cases with examples.}

\subsection{Android App Dataset}
We prepared a dataset of 14 frequently downloaded apps for use in our evaluations.
They were randomly selected from 312 apps evaluated in~\cite{fok_large-scale_chi22}, covering nine common app categories: Education, Entertainment, Events, Food \& Drinks, Housing, Music, Shopping, Sports, and Travel.
During the selection process, whenever a selected app was no longer available from the Google Play store, we excluded it and selected an alternative app with \add{a} similar number of downloads.
We encountered two such apps.
We also excluded \edit{apps}{one app} that \edit{were}{was} completely inoperable with screen readers, as our tool currently only covers elements reachable using a screen reader.
We believe a combination of heuristics and LLM analysis can be used to \edit{identify such instances}{audit such apps}, \add{taking inspirations from \cite{chiou2023bagel, Salehnamadi2023groundhog}}, and we leave this case for future research.

\subsection{Expert Profiles}
\begin{table*}[ht]
  \caption{Expert Participant Profiles.}
  \label{tab:experts}
  
  \renewcommand{\arraystretch}{1.2}
    \begin{tabular}{c p{6.5cm} c p{2.7cm} p{5.5cm} }
        \toprule
        \textbf{Id} & \textbf{Main Accessibility Experience} & \textbf{YoE} & \textbf{Current Role} & \textbf{AT Experience} \\
        \midrule
        \rowcolor{gray!10} E1 & Developer of an iOS app designed for people who are blind or low vision at a  software company. & 8 & PhD Student & VoiceOver, iOS Accessibility Inspector \\
        E2 & PDF Document accessibility research at a  software company and at a university. & 2.5 & PhD Student & NVDA, VoiceOver, AxesPDF, WAVE, PDF checkers \\
        \rowcolor{gray!10} E3 & Technology specialist, web developer at a university's accessible technology center. & 24 & Director & JAWS, NVDA \\ 
        E4 & Internal accessibility consulting and usability testing at a software company's accessibility team. Teaches programmatic access and designing for people with disabilities at a university. & 24 & Professor & Narrator, JAWS, company's internal accessibility testing tools \\
        \rowcolor{gray!10} E5 & Research with screen reader users, including using AR for navigation and non-visual games at a software company. & 12 & Researcher, Software Engineer & Screen Readers, Augmentative and Alternative Communication (AAC) devices \\
        E6 & Mobile app accessibility consultant, screen reader user (blind). & 25 & Program Manager & VoiceOver, TalkBack \\ 
        \bottomrule
    \end{tabular}
\end{table*}

We recruited six accessibility experts (E1--E6) with 
an average of 15.9 ($sd = 8.9$) years of experience (YoE) directly related to designing, developing, researching, or using access technology (AT) for people who are blind or low vision.
All participants are currently working in accessibility-related roles and self-described as very familiar with screen readers.
E1--E5 all have accessibility research or development experience at major software companies and in academic settings.
E6 is blind and uses screen readers.
He consults and coaches mobile app developers to help make their apps accessible.
Table~\ref{tab:experts} details the profile of each expert.
\add{We provided study information to each participant as part of our invitation, including its description, duration, and compensation.
All participants voluntarily chose to participate in the study.}

\subsection{Procedure} \label{sec:expert-proc}
We conducted one evaluation session with each participant\add{, which lasted between 60 and 90 minutes.
Each participant was provided with a 50 USD gift card as compensation.}
Each session consisted of three parts: an introduction, \add{a few} evaluation tasks\add{, and a semi-structured interview}.
Time permitting, we presented up to four randomly chosen mobile app screens from our dataset.

First, we introduced the basics of TalkBack and common screen reader accessibility guidelines adapted from the Appt Guidelines for mobile apps~\cite{appt_guidelines} as a reminder.
We also demonstrated the use of Accessibility Scanner and \toolName's report on an app outside of our dataset. 
The introduction took between 10 and 15 minutes.

Next, participants performed independent evaluations of each mobile app screen. 
Participants were asked to use TalkBack to interact with the screen and identify any \add{screen reader} accessibility \edit{failures}{errors} and take brief notes in a blank study worksheet.
\add{Participants were also offered a report from the Android Accessibility Scanner~\cite{a11y_scanner}.}
After \edit{interaction}{the assessment}, we asked each participant to explain all observed \edit{failures}{errors} verbally.
We then presented the participant with \toolName's report for the same screen, which was not offered during their evaluation.
Participants were asked to note any missing issues or inaccurate descriptions.
\edit{This process was applied to each screen, with}{There were} no time constraints.
The evaluations took between \edit{35}{40} and \edit{45}{60} minutes.

Finally, we conducted a semi-structured interview to gather additional feedback.
We used open-ended questions to elicit feedback on the tool’s usability, coverage, and potential impact on accessibility practices.
The interview took between 10 and 15 minutes.

We conducted the sessions with E1--E5 in-person and with E6 remotely.
The above procedure was modified for E6 with three changes:
(1)~We did not explain the guidelines and use of screen readers as E6 was blind and a screen reader user.
(2)~Transcripts and descriptions of the app screens were made available to E6 prior to the study.
(3) During the study, the researcher first verbally described the screen then went through each transcript entry and described its corresponding UI element when necessary, and asked the participant for an evaluation.
E6 was then read the \toolName's analysis and \edit{evaluated}{asked to evaluate} the report entry.
This process repeated for each app.

\subsection{Analysis Overview}
We collected audio recordings of each study session and each participant's study worksheet.
Participants analyzed between 1 and 4 screens in each session, with an average of 2.3 screens.
In total, participants covered 14 unique screens, each in a different app, containing 306 UI elements.
We \edit{analyzed}{processed} the collected data in two parts: (1)~accessibility error analysis of all UI elements \add{and construction of the accessibility error dataset} and (2)~analysis of qualitative data gathered during interviews.

\begin{table}[ht]
    \centering
    \caption{Accessibility error categories identified by experts and their corresponding WCAG 2.1 success criteria.}
    \label{tab:errors}
    \renewcommand{\arraystretch}{1.2}
    \begin{tabular}{p{2.45cm} p{4cm} c}
        \toprule
        \textbf{Error Category}        & \textbf{Success Criteria}    & \textbf{Count}   \\
        \midrule
        \rowcolor{gray!10}Missing Label         & 1.1.1 Non-text Content & 39  \\
        Label Quality         & 2.4.4 Link Purpose (In Context)\newline 2.4.6 Headings and Labels\newline 4.1.2 Name, Role, Value  & 42  \\
        \rowcolor{gray!10}Structure\;\& Grouping & 1.3.1 Info and Relationships\newline 1.3.2 Meaningful Sequence\newline 2.4.3 Focus Order\newline 3.2.3 Consistent Navigation   & 54  \\
        Heading               & 2.4.2 Page Titled\newline 2.4.10 Section Headings   & 16  \\
        \rowcolor{gray!10}Functionality         & 2.1.1 Keyboard\newline 3.2.1 On Focus\newline 3.2.2 On Input\newline 4.1.3 Status Messages & 12  \\
        (No Error)              & /   & 143 \\
        \midrule
        \rowcolor{gray!10}\textbf{Total}                   & /     & 306 \\
        \bottomrule
    \end{tabular}
\end{table}

\subsection{Accessibility Error Analysis and Dataset Construction}
To obtain an objective accessibility error determination of all UI elements, two researchers independently analyzed participant analyses of all screens and assigned relevant transcripts or worksheet notes to their relevant UI elements.
The researchers then individually summarized the accessibility error(s) for each UI element based on these transcripts and notes, while cross-referencing WCAG~2.1~\cite{w3c_wcag2.1} and MagentaA11y's native app accessibility guidelines~\cite{magentaa11y}.
Finally, the researchers combined their analysis and discussed any inconsistencies, resulting in a final list of accessibility errors for all UI elements.

Our analysis revealed 163 accessibility errors across the 14 screens evaluated.
To further understand the types of errors encountered, we classified the identified errors into five categories: (1)~\textit{Missing Label}, (2)~\textit{Label Quality}, (3) \textit{Structure \& Grouping}, (4) \textit{Heading}, (5) \textit{Functionality}.
These classifications are guided by expert comments and the WCAG 2.1 accessibility guidelines.
We identified the most relevant success criteria observed within each error category and their number of occurrence as a reference, detailed in Table~\ref{tab:errors}.

\begin{figure*}
\noindent\begin{minipage}{\linewidth}
\newcommand{\form}[1]{\textcolor{AddColor}{#1}}
\newcolumntype{Z}{>{\RaggedRight\arraybackslash\form}p{2.87cm}}
    \centering
    \addImg{\includegraphics[width=0.99\textwidth]{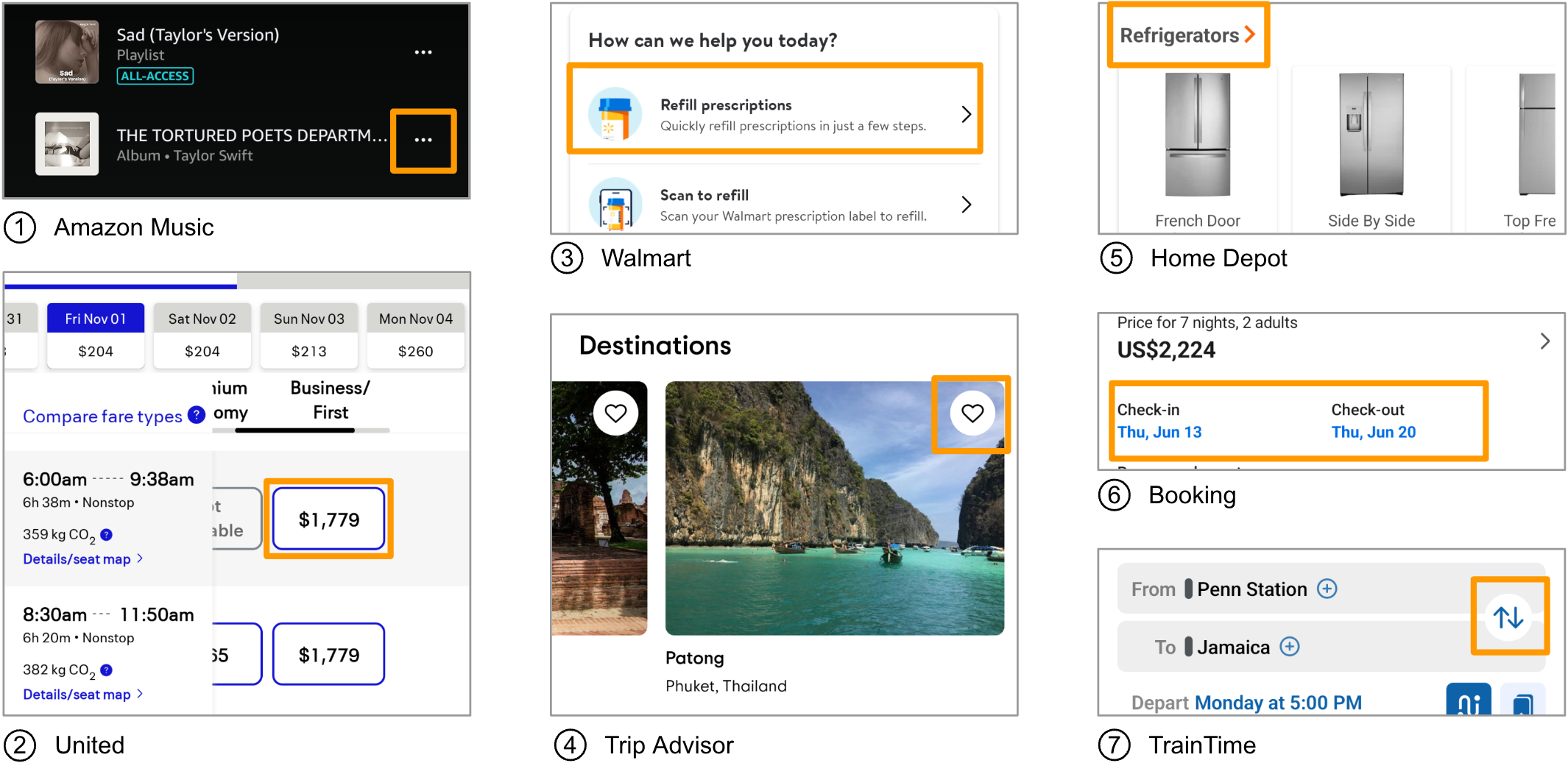}}

\vspace*{12pt}

    \centering
    \small
    \renewcommand{\arraystretch}{1.5}
    \selectfont
    \arrayrulecolor{AddColor}
    \begin{tabular}{>{\color{AddColor}}c ZZZZZ >{\color{AddColor}}c}
        \toprule
        \textbf{\#} & \textbf{App \& UI Element} & \textbf{TalkBack Transcript} & \textbf{Expert Analysis} & \textbf{\toolName} & \textbf{Accessibility Scanner} & \textbf{Type} \\
        \midrule
        \rowcolor{gray!10}\circled{1} & Amazon Music,\newline ``View More" Button & res/drawable/ic\_action\_more.xml (\textit{Repeated for consecutive swipes.}) & (1)~Improper content description for Image Button. (2)~Duplicated button. & (1)~Using internal identifiers as labels. (2)~Repeating or redundant labels. & Multiple clickable items share this location on the screen. & \Gape[0pt][2pt]{\makecell[tc]{LQ \\ SG}} \\
        \circled{2} & United,\newline Price Selection & \$1,779. Double-tap to activate. & Table element should announce column and row headers. & The price without context is unclear. & Multiple items have the same description. & Head \\
        \rowcolor{gray!10}\circled{3} & Walmart,\newline "Refill Prescriptions" & ...Quickly refill prescriptions in just a few steps 2 Period Opens in a browser... & Unnecessary and confusing text ``2~Period'' in label. & The statement \textquotesingle 2~Period\textquotesingle{} in the speech output is confusing. & n/a & LQ \\
        \circled{4} & Trip Advisor,\newline ``Save to Trip" Button & Save to a trip. Double-tap to Save to a trip. Double-tap and hold to Save to a trip. & (1)~Relationship between button and the trip to be saved unclear. (2)~Repeated speech unnecessary. & (1)~Button label unnecessarily repeats the action hint.\newline (2)~Multiple labels are identical with no unique context. & n/a & \makecell[tc]{SG \\ LQ} \\
        \rowcolor{gray!10}\circled{5} & Home Depot,\newline ``Refrigerators'' Heading & Refrigerators. Double-tap to activate. & (1)~No heading indication. (2)~Activation target unclear. & Elements ... seem to be related and should be grouped. & n/a & \makecell[tc]{Head \\ Func} \\
        \circled{6} & Booking,\newline ``Check-in,'' ``Check-out'' Options & \,-- Check-in \newline \,-- Check-out \newline \,-- Thu, Jun 13 \newline \,-- Thu, Jun 20 & Grouping and ordering issue. & n/a & n/a & SG \\
        \rowcolor{gray!10}\circled{7} & TrainTime,\newline ``Switch'' Button. & (\textit{Not announced.}) & The ``Switch From/To'' button is not focusable. & n/a & n/a & Func \\
        \bottomrule
    \end{tabular}

    \vspace*{4pt}
    \captionof{figure}{\add{Accessibility errors identified by \toolName and Accessibility Scanner. Axe~\cite{axe} did not identify any of the errors listed. Abbreviations: LQ = Label Quality. SG = Structure \& Grouping. Head = Heading. Func = Functionality.}}
    \label{fig:examples}

\end{minipage}
\end{figure*}

\add{\subsection{Illustrative Examples}} \label{sec:illustrative-examples}
\add{We show examples to illustrate the accessibility errors identified by the experts and the outputs from \toolName and Accessibility Scanner.
Figure~\ref{fig:examples} shows seven exemplar UI element TalkBack transcripts, expert-identified accessibility errors, and outputs from the tools.
Each example may show more than one error, because (1)~a UI element may trigger multiple TalkBack announcements or (2)~a single announcement many contain more than one error.
}

\add{Specifically, we consider \toolName's outputs for examples \circled{1}\,--\,\circled{4} to be consistent with expert analyses, while its outputs for examples \circled{5}\,--\,\circled{7} are inconsistent.
Next, in Sections~\ref{sec:neglect} and \ref{sec:expert-limitations}, we discuss these examples in the context of expert comments.
}

\subsection{Qualitative Feedback from Experts} \label{sec:expert-feedback}
\add{We received expert feedback during their evaluation tasks and after the tasks during the interview.
While analyzing feedback,}
we followed guidelines for thematic analysis~\cite{braun2006thematic} which involved an inductive and deductive coding process.
\add{Experts in general found \toolName to be (1)~\textit{effective and consistent}, (2)~\textit{easy to understand and to use}, (3)~\textit{capable of uncovering neglected errors}, but also identified its (4)~\textit{limitations}.}

\subsubsection{Effectiveness and Consistency}
Experts found \toolName consistently and effectively identified major accessibility issues within mobile applications.
    \begin{lstlisting}
        <@It definitely did a better job than the current tool because I think the current tool would like identify some things and then randomly miss it for a very similar issue. This was a lot more consistent. (E1)@>
        <@I see a ton of benefit in the fact that it points out a whole bunch of different things to think about and it makes it really easy too. (E1)@>
        <@I think it picked up most of what I picked up. (E2)@>
        <@But it does seem to be pulling up all the main points that I did, and I can't think of what else it would say. That's great to know. But it's interesting. (E3)@>
        <@The fact that this is a tool is extremely helpful. (E6)@>
    \end{lstlisting}

\subsubsection{\add{Ease of Understanding and Use}}
Experts found \toolName's interface and generated reports accessible and easy to comprehend, which makes it user-friendly for developers who may not be deeply familiar with accessibility.
    \begin{lstlisting}
        <@So I think the two views of the UI and where they are in the UI, but also just being able to go through a list is, is really helpful. Also, you have an accessible view with the list, which is always good. (E1)@>
        <@The issues are in a human readable format. So I think over time, you would really like there, it's easy to comprehend. (E4)@>
    \end{lstlisting}

\subsubsection{Uncovering Neglected Errors} \label{sec:neglect}
\toolName identified accessibility errors that experts considered to be hard to identify or errors that experts initially missed.
\add{These cases require the model to understand interface structure or identify low-quality labels.}

\begin{enumerate}
    \item \add{\textbf{Structure Understanding:} In example~\circled{1} of Table~\ref{tab:experts}, \toolName impressed E1 on its ability to understand navigation. In example~\circled{2}, E2 appreciated ScreenAudit’s ability to identify the context.}
    \begin{lstlisting}
      <@\circled{1} I think search results are a very special case in the sense that it's one of the more difficult navigation things and it would be very difficult to pick that up. So, like, there's something about navigation, which I would actually say was probably a huge problem. But it did. (E1)@>
      <@\circled{2} And since this is not a part of any header, then maybe it should just simply talk about it. But since it's a part of the column, it should identify it in the form of a table. So that it identified is really great. (E2)@>
    \end{lstlisting}
    \item \add{\textbf{Identifying Low-Quality Labels:} In example~\circled{3}, \toolName identified an extra announcement ``2 Period'', which was initially missed by E5 during the study. In example~\circled{4}, E3 initially did not recognize the repeated announcements that were detected by \toolName.}
    \begin{lstlisting}
      <@\circled{4} Quite nicely. I wasn't necessarily... Was I hearing all the ``double-tap'' and ``double-tap and hold'' stuff? Oh, I think I was just tuning out. It says they are identical. That's true. And the corresponding location it's associated with is what I was saying. (E3)@>
    \end{lstlisting}
\end{enumerate}

\subsubsection{\add{Limitations}} \label{sec:expert-limitations}
\add{Experts identified three main limitations in their feedback during and after the studies: (1) \textit{inaccurate results}, (2) \textit{not fully understanding context}, and (3) \textit{unable to test app functionality}.}
\begin{enumerate}
    \item \add{\textbf{Inaccurate Results:} Sometimes, \toolName identifies an error, but its explanation does not directly address the root cause. In example~\circled{5}, \toolName identified a structural issue but did not suggest add a heading. E3 noted this and commented that, while the output was incorrect, it could provoke developers to think about their design.}
    \item \add{\textbf{Not Fully Understanding Context:} Experts identified the main limitation of \toolName to be its limited ability in understanding context surrounding elements (e.g., example~\circled{6}), while noting that all current tools have problem in this aspect.}
    \begin{lstlisting}
      <@\circled{6} I think the only thing was like sort of missing things, which I think were usually because it was more context based. Which I also understand isn't really, um, isn't really something there's even like a set standard for. (E1)@>
    \end{lstlisting}
    \item \add{\textbf{Unable to Test App Functionality:} Another limitation experts identified was that \toolName was not able to directly test the functionality on the UI. In example~\circled{7}, the switch button could not be focused by the screen reader and was ignored. E5 identified this in his testing and pointed out:}
    \begin{lstlisting}
      <@\circled{7} It's not catching interaction things though, and I think that's a really critical problem. None of the tools do. (E5)@>
    \end{lstlisting}
\end{enumerate}

\section{\edit{Prompt}{Performance} Evaluation} \label{sec:prompt-eval}
    \addDiscuss{[Revision Note] We added three analyses: comparison with existing checkers (Section~\ref{sec:compare-tools}), output consistency evaluation (Section~\ref{sec:compare-consistency}), and experiment with different LLMs (Section~\ref{sec:compare-llms}). We also added statistical tests to existing analyses to make the claims more robust.}

\add{In this section, we evaluate the performance of \toolName and provide insights into its prompt design. We conduct our analyses in five respects:
(1)~Comparing the performance of \toolName and two current accessibility checkers.
(2)~Comparing the performance of \toolName's five prompts.
(3)~Understanding whether \toolName's outputs are consistent across multiple runs.
(4)~Examining \toolName's performance by each accessibility error category.
(5)~Comparing the performance of currently available LLMs and understand whether these prompting techniques can be transferred to other LLMs.}

\add{\subsection{Experiment Method}}

\add{All our experiments were performed on the accessibility error dataset created in the expert study.}
\add{Unless otherwise specified, we used OpenAI's GPT-4o, 2024-08-06 version.}

\add{For each analysis,} two of the researchers independently compared \edit{the \toolName's}{a tool's} output for each UI element to \del{that of }the \add{corresponding} ground truth label collected in the expert study.
Each researcher assigned \textit{``consistent''} or \textit{``inconsistent''} to \add{each of} the tool's outputs\add{, which we refer to as \textit{Correctness}}.
\edit{The}{An} output was found to be \textit{inconsistent} if it did not identify an issue, identified a different issue, or provided an incorrect explanation.
After this process was done, the researchers discussed until an agreement \edit{is}{was} reached on all elements.

\add{\subsection{Analyses and Results}}

\subsubsection{\add{Comparison of Accessibility Evaluation Tools}} \label{sec:compare-tools}

\add{We compared \toolName against two widely-used accessibility checkers for Android: Google's Accessibility Scanner~\cite{a11y_scanner} and Deque's Axe~\cite{axe}.
Axe was not able to run on one screen in the dataset (United's booking screen), therefore that screen was excluded for Axe.}

\add{Table~\ref{tab:tool-comparison} shows precision, recall, and F1 score for each of the tools tested.
An analysis of variance based on mixed logistic regression~\cite{robert1984random, gilmour1985analysis} indicated a statistically significant effect of \textit{Tool} on \textit{Correctness}, $\chi^2(2, N\mathord{=}2142)=170.1$, $p<.001$.
Post hoc pairwise comparisons, corrected with Holm’s sequential Bonferroni procedure~\cite{holm_bonferroni}, indicated that \toolName vs. Accessibility Scanner ($Z=8.08, p<.001$), \toolName vs. Axe ($Z=11.02, p<.001$), and Accessibility Scanner vs. Axe ($Z=4.11, p<.001$) were significantly different.
The Accessibility Scanner and Axe's lower recall agrees with prior research that current automated checkers only detect a small subset of accessibility errors~\cite{Carvalho_2018_a11yproblems, mateus2020eval}.
ScreenAudit is able to cover a much higher percentage of errors with a modest reduction in precision.}

\begin{table}[ht]
    \caption{\add{Performance metrics for \toolName, Accessibility Scanner, and Axe.}}
    \label{tab:tool-comparison}
    \begin{minipage}{\columnwidth}
        \begin{center}
        \begin{tabular}{>{\color{AddColor}}l >{\color{AddColor}}c >{\color{AddColor}}c >{\color{AddColor}}c}
            \arrayrulecolor{AddColor}
            \toprule
            \textbf{Tool} & \textbf{Precision} & \textbf{Recall} & \textbf{F1} \\
            \midrule
            \toolName* & .713 & \textbf{.622} & \textbf{.664} \\
            Accessibility Scanner & .729 & .313 & .438 \\
            Axe & \textbf{.812} & .171 & .283 \\
            \bottomrule
        \end{tabular}
        \end{center}

        \bigskip
        \footnotesize \add{ *\; \toolName's statistics are averages of the five prompts evaluated in Section~\ref{sec:prompt-comparison}.}
    \end{minipage}
\end{table}

\subsubsection{\add{Comparison of Prompts}} \label{sec:prompt-comparison}
We tested the following five prompts \edit{based on}{according to} their structures outlined in Section~\ref{sec:prompt}:
(1) \textit{Base}: a baseline condition with only an introduction and task instructions.
(2) \textit{General}: \textit{Base} prompt with added general accessibility guidelines.
(3) \textit{Contextual}: \textit{Base} prompt with added instructions about looking for contextual cues.
(4) \textit{General\_Contextual}: Combined prompt with \textit{General} accessibility guidelines and \textit{Contextual} instructions.
(5) \textit{WCAG\_Contextual}: Combined prompt with \textit{WCAG} accessibility guidelines and \textit{Contextual} instructions.
\del{For details on each condition, please refer back to Section~\ref{sec:prompt}.}

\begin{table*}[ht]
    \centering
    \caption{\edit{Comparison of}{\toolName's} prompt performance on the accessibility error dataset collected in the expert study.}
    \label{tab:prompt-results}
    \begin{tabular}{lcccccc}
        \toprule
        \edit{Model}{\textbf{Prompt}} & \textbf{Introduction} & \textbf{Accessibility} & \textbf{Instruction} & \textbf{Precision} & \textbf{Recall} & \textbf{F1} \\
        \midrule
        \textit{Base} & Yes & / & Basic & \textbf{.797} & .577 & .669 \\
        \textit{General} & Yes & General & Basic & .696 & .589 & .638 \\
        \textit{Contextual} & Yes & / & Contextual & .667 & .650 & .658 \\
        \textit{General\_Contextual} & Yes & General & Contextual & .708 & \textbf{.699} & \textbf{.704} \\
        \textit{WCAG\_Contextual} & Yes & WCAG & Contextual & .698 & .595 & .642 \\
        \bottomrule
    \end{tabular}
\end{table*}

Table~\ref{tab:prompt-results} shows the evaluation results for the five prompts.
\add{An analysis of variance based on mixed logistic regression~\cite{robert1984random, gilmour1985analysis} indicated a statistically significant effect of \textit{Prompt} on \textit{Correctness}, $\chi^2(4, N\mathord{=}1530)=18.66$, $p<.001$.
Post hoc pairwise comparisons, corrected with Holm’s sequential Bonferroni procedure~\cite{holm_bonferroni}, indicated that \textit{Base} vs. \textit{Contextual} ($Z=3.28, p=.010$) and \textit{Base} vs. \textit{General\_Contextual} ($Z=4.02, p<.001$) were significantly different.
}

\edit{The}{Our results show that \toolName's} prompt with general accessibility guidelines and contextual instructions has the highest recall and F1 score.
It appears that the \edit{existence of each of these components individually}{addition of either general accessibility guidance or contextual instruction alone} does not \edit{contribute to an improvement in}{improve} model performance \add{over the \textit{Base} prompt}.
However, a combination of both improves the performance\del{ over the \textit{Base} condition}.
Interestingly, \del{simply }adding the WCAG guidelines \edit{to the prompt}{instead of general accessibility guidance} does not yield better performance \add{compared to the \textit{Base} prompt}.
\edit{This is likely because}{One possible reason is that} the WCAG guidelines are \del{overly }abstract and technical, making it difficult for the LLM to effectively utilize.
\add{We further discuss this finding in Section~\ref{sec:heuristic-eval}.}

\subsubsection{\add{Consistency}}  \label{sec:compare-consistency}
\add{We ran the \textit{General\_Contextual} prompt four additional times on the accessibility error dataset to evaluate \toolName's output consistency.
We performed an analysis of variance based on mixed logistic regression~\cite{robert1984random, gilmour1985analysis} on the five runs and found no detectable difference in \textit{Correctness}, $\chi^2(4, N\mathord{=}1530)=3.24$, $p=.52$.
Given the amount of data analyzed, any meaningful difference in output would likely have been detected; therefore, we found the model output to be consistent across the evaluated runs.
On average, the model achieved a precision of .723 ($SD=.038$) and a recall of .692 ($SD=.023$), with an overall F1 score of .707.}


\subsubsection{\add{\toolName Performance by Error Category}}

\begin{figure}[ht]
    \centering
    \addImg{\includegraphics[width=0.4\textwidth]{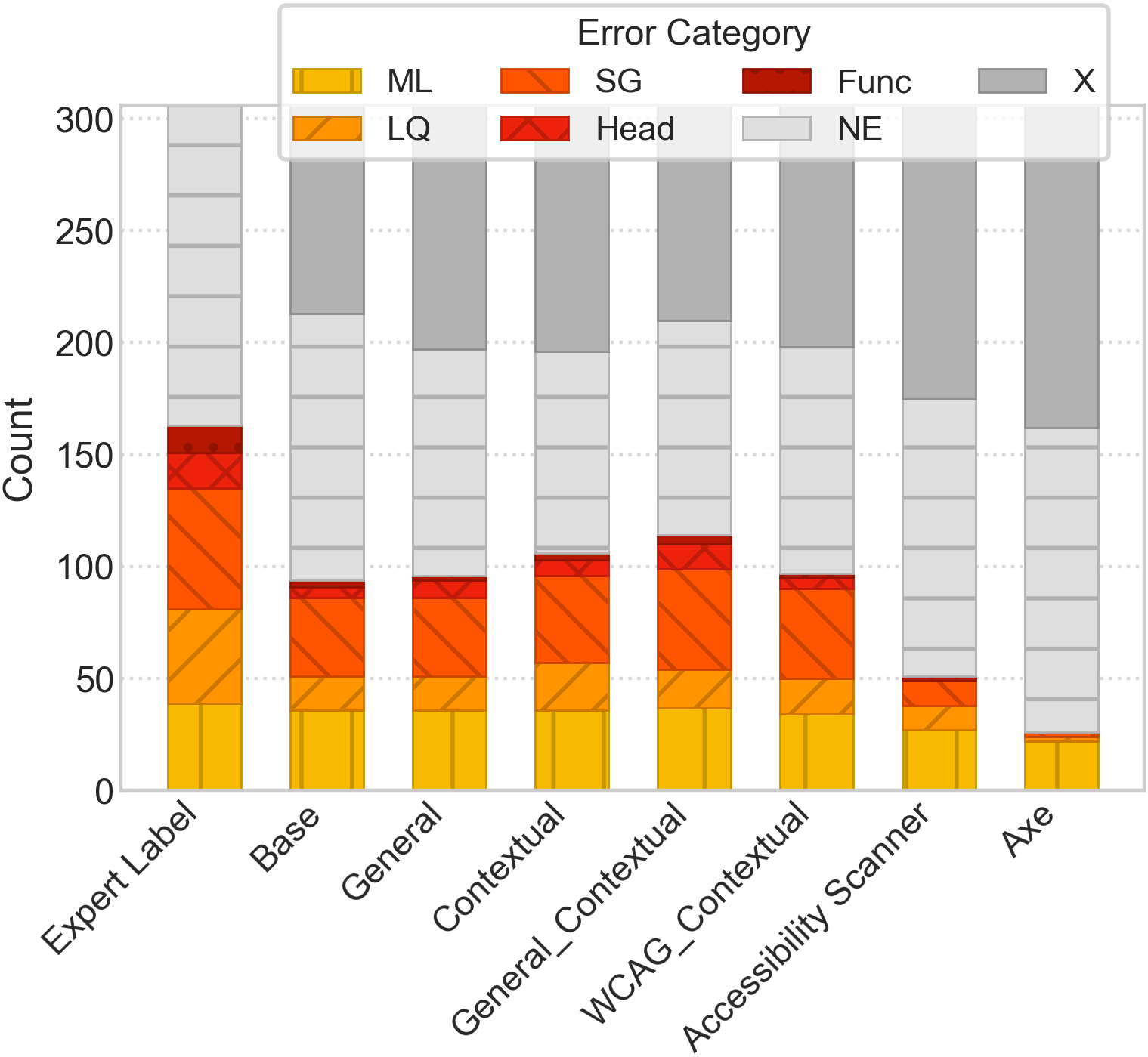}}
    \caption{\add{Stacked bar plot for each prompt or tool's performance by error category, with expert labels on the left as reference. Counts for ML, LQ, SG, Head, Func represent the number of true positives. Count for NE represents the number of true negatives. Count for X represents the number of classification errors.}}
    \label{fig:perf_stacked_count}
\end{figure}

We further break down each prompt's \add{and each accessibility checker's} performance by \del{the }accessibility error category in Table~\ref{tab:prompt-by-category}
\add{and visualize their true positives, true negatives, and classification errors in Figure~\ref{fig:perf_stacked_count}}.
As expected, Missing Label errors are \add{relatively} easy to detect by automated checkers and therefore have high\add{er} detection rates across all prompts \add{and checkers}.
\add{Nevertheless, \toolName outperforms rule-based checkers in all accessibility error categories.
Comparing those elements with no errors, rule-based checkers have lower false positive rates than \toolName.
Future research can investigate whether rule-based checkers can support LLMs, for example through Retrieval Augmented Generation (RAG), to lower false positive rates.}

\add{For \toolName,} it appears that general accessibility guidelines improved the model's performance in detecting errors with Headings, while contextual instructions improved the model's performance in detecting errors with Label Quality and Structure \& Grouping.
The lower performance of all prompts in Label Quality and Functionality is likely because \del{the }\toolName does not currently explore the interactivity or functionality of elements.
Therefore, it becomes more difficult for \toolName to detect such errors as many are context-dependent.

\begin{table}[ht]
\centering
\caption{Break down of prompt \add{or tool's} accuracy \add{(in percentages \%)} by accessibility error category.}
\semismall
\label{tab:prompt-by-category}
\begin{minipage}{\columnwidth}
    \begin{center}
    
    \begin{tabular}{lcccccc}
    \toprule
    \textbf{Prompt\add{/Tool}} & \textbf{ML} & \textbf{LQ} & \textbf{SG} & \textbf{Head} & \textbf{Func} & \textbf{NE} \\
    \midrule
    \textit{Base}                            & 92.3 & 35.7 & 64.8 & 31.2 & 25.0 & 83.2 \\
    \textit{General}                         & 92.3 & 35.7 & 64.8 & 50.0 & 16.7 & 70.6 \\
    \textit{Contextual}                      & 92.3 & \textbf{50.0} & 72.2 & 43.8 & 25.0 & 62.9 \\
    \edit{Contextual\_General}{\textit{General\_Contextual}} & \textbf{94.9} & 40.5 & \textbf{83.3} & \textbf{68.8} & \textbf{33.3} & 67.1 \\
    \edit{Contextual\_WCAG}{\textit{WCAG\_Contextual}} & 87.2 & 38.1 & 74.1 & 31.2 & 16.7 & 70.6 \\
    \midrule
    \add{A11y Scanner }    & \add{69.2} & \add{26.2} & \add{20.4} & \add{0.0} & \add{16.7} & \add{86.7} \\
    \add{Axe }                      & \add{56.4} & \add{5.3} & \add{4.0} & \add{0.0} & \add{0.0} & \add{\textbf{94.9}} \\
    \bottomrule
    \end{tabular}
    \end{center}

    \bigskip
        \footnotesize \add{ \emph{Abbreviations:} $\boldsymbol{\cdot}$ ML = Missing Label\quad$\boldsymbol{\cdot}$ LQ = Label Quality\quad$\boldsymbol{\cdot}$ SG = Structure \& Grouping\quad$\boldsymbol{\cdot}$ Head = Heading\quad$\boldsymbol{\cdot}$ Func = Functionality\quad$\boldsymbol{\cdot}$ NE = No Error}
\end{minipage}

\end{table}

\subsubsection{\add{Comparison of LLMs}}  \label{sec:compare-llms}

\add{We compared the performance of GPT\nobreakdash-4o, the LLM that we used in the above experiments, against other state-of-the-art LLMs: OpenAI o1~\cite{openai-o1}, Anthropic's Claude~3.5 Sonnet~\cite{claude35sonnet}, Google's Gemini~1.5 Pro~\cite{gemini}, and Meta's Llama 3.1 405B Instruct~\cite{dubey2024llama3herdmodels}.
Table~\ref{tab:llm-perf} shows each model's recall, precision, and F1 score when using the \textit{Base} and the \textit{General\_Contextual} prompts on the accessibility error dataset.}
\add{OpenAI o1 and Gemini both achieved better performance with the \textit{General\_Contextual} prompt over the \textit{Base} prompt.
Claude's performance change was negligible, while Llama 3.1's performance decreased.
These variations among LLMs may result from the differences in model steerability~\cite{chang2024measuring}, or be influenced by our experiment process, which focused on GPT-4o.
}

\begin{table*}[ht]
    \centering
    \setlength{\tabcolsep}{4pt}
    \caption{\add{Comparison of LLM performance on the accessibility error dataset with the Base and General\_Contextual (GC) prompts. $\Delta=$ change.}}
    \label{tab:llm-perf}
\begin{tabular}{
    >{\color{AddColor}}l 
    >{\color{AddColor}}c 
    >{\color{AddColor}}c 
    >{\color{AddColor}}c 
    c    
    >{\color{AddColor}}c 
    >{\color{AddColor}}c 
    >{\color{AddColor}}c 
    c    
    >{\color{AddColor}}c 
    >{\color{AddColor}}c 
    >{\color{AddColor}}c
}
    \arrayrulecolor{AddColor}
    \toprule
    & \multicolumn{3}{c}{\add{\textbf{Precision}}} 
    & 
    & \multicolumn{3}{c}{\add{\textbf{Recall}}} 
    & 
    & \multicolumn{3}{c}{\add{\textbf{F1}}} \\
    \cmidrule(lr){2-4} \cmidrule(lr){6-8} \cmidrule(lr){10-12}
    \textbf{Model} 
    & \textbf{Base} & \textbf{GC} & \(\boldsymbol{\Delta}\)
    & 
    & \textbf{Base} & \textbf{GC} & \(\boldsymbol{\Delta}\)
    & 
    & \textbf{Base} & \textbf{GC} & \(\boldsymbol{\Delta}\) \\
    \midrule
        GPT-4o (2024-08-06) & .797 & .723 & -.074 & & .577 & .692 & +.115 & & .669 & .707 & +.038 \\
        OpenAI o1 (2024-09-12) & .606 & .675 & +.069 & & .577 & .679 & +.102 & & .591 & .677 & +.086 \\
        Claude 3.5 Sonnet (2024-10-22) & .714 & .724 & +.010 & & .337 & .337 & .000 & & .458 & .460 & +.002 \\
        Gemini 1.5 Pro-002 & .582 & .617 & +.035 & & .607 & .693 & +.086 & & .595 & .653 & +.058 \\
        Llama 3.1 405B Instruct & .686 & .644 & -.042 & & .429 & .380 & -.049 & & .528 & .478 & -.050 \\
        \bottomrule
    \end{tabular}
\end{table*}

\section{Discussion}
    \addDiscuss{[Revision Note] We added two additional discussion points on how \toolName relates to and adds to the existing body of knowledge, specifically on heuristic evaluation and simulated user testing.}

Our results show that \toolName achieves significant coverage of screen reader accessibility errors in mobile apps, with a recall of 69.2\%.
\add{Our fundamental insight is to simulate \textit{how} a user interacts with apps using a screen reader and directly analyze that experience.}
We explored the effectiveness of leveraging LLMs for \edit{identify}{performing these analyses} and providing explanations for \add{the detected} accessibility errors\del{ in mobile apps}, revealing promising research directions for future accessibility evaluation tools.
\edit{Our expert evaluation demonstrated the effectiveness and usefulness of \toolName and identified areas for future research and development.}{In this section, we discuss how \toolName complements and has the potential of integrating rule-based checkers, how some of our technique relates to heuristic evaluation, and the potential of simulating user testing with LLMs.}

\subsection{\toolName \edit{Complementary to Existing Accessibility Evaluation Tools}{Complements Existing Accessibility Checkers in Creating Accessible Apps}}
Our goal in developing \toolName is not to replace current rule-based accessibility checkers with an LLM-based checker.
Instead, we \del{strongly }recognize the speed, cost-effectiveness, and consistency of these rule-based checkers.
We envision integrating the techniques of \toolName into open-source accessibility checkers \add{to combine the strengths of rule-based checkers and those of \toolName.}
\add{We also hope} to explore utilizing \edit{these results}{the results from rule-based checkers~\cite{axe, google-atf, apple_use_2023} and crawler-based checkers~\cite{wave, Salehnamadi2023groundhog, chiou2023bagel}} as part of \edit{the input to LLM evaluation}{LLM input}\del{.
Existing rule-based checkers can potentially also be used} to \add{provide additional context or} \edit{double-check}{to validate} the correctness of LLM evaluations and iteratively build a final assessment.

\add{Although we primarily target screen reader accessibility errors, the same accessibility representations support a range of access technologies, such as switch access, keyboard navigation, and voice control.
Given that developers have the most comprehensive understanding of their app's UI layouts and visual design intent, they are best positioned to provide and ensure the accuracy of accessibility metadata.
By directly supporting developers, we aim to promote the creation of more accessible apps.
}

\add{\subsection{LLMs and Heuristic Evaluation}} \label{sec:heuristic-eval}
In our experiments, we found that providing general accessibility guidance to the LLM yielded better results than providing it with WCAG guidelines.
Although this appears counter-intuitive, there has been research that concluded WCAG guidelines only covered a subset of all accessibility issues encountered by users~\cite{power2012_guidelines, casare2016usability}.
We also draw parallel to Nielson's research on usability heuristics~\cite{nielson1990heuristic}, where these heuristics have been shown to effectively guide usability evaluation and approximate the results from user studies.
We differ from usability heuristics by tailoring our heuristics to accessibility issues and using LLMs instead of experts or specialists.
Nevertheless, our results suggest that adopting a heuristic evaluation approach with LLMs may be beneficial compared to exhaustively listing all possible error types.

\add{\subsection{Towards Simulated User Testing with LLMs}}
\add{Research has studied LLM-supported user behavior simulations for web-marketing campaigns~\cite{kasuga2024cxsimulator} and smartwatch interfaces~\cite{xiang2024simuser}.
However, while LLMs can hypothesize how an interface may respond to users with different abilities, they are unable to test the various ways of interacting with an interface, such as with screen reader, eye tracking, or switch access.
Without such information, LLMs can be prone to hallucinations~\cite{xiang2024simuser}.
\toolName grounds LLM analysis of accessibility by directly interacting with an app using a screen reader.
Although unlikely to eliminate \textit{all} hallucinations, the approach of interacting with access technologies and analyzing their feedback can be adapted to simulate diverse abilities from users.
Contextual information from such interactions should reduce the likelihood of undesired response from the LLMs, enabling fast and cost-effective feedback for designers and developers.
}

\section{Limitations and Future Work}
    \add{Although \toolName demonstrates promising capabilities in detecting screen reader accessibility errors, we also identify its limitations and suggest directions for future research.}

\subsection{UI Element Coverage}
One of the main constraint\add{s} of \toolName is that it utilizes TalkBack's core pipeline to traverse screen reader focusable elements on an app screen.
This dependency limits \toolName's ability to explore and detect elements that are actionable but unfocusable to screen readers.
We expect future developments to \add{additionally} utilize structured UI information, such as the view hierarchy from the Accessibility API, or \edit{by extracting}{extracted UI} structures from screenshots~\cite{wu2021screen}\add{, so that all elements perceivable by sighted users will be examined}.

\subsection{Acquiring Additional Context}
As experts pointed \add{out} in their comments, context understanding is an important yet unsolved problem of all accessibility checkers.
Although our adoption of LLMs provides a first step \edit{of}{towards} addressing this challenge, additional research is still needed. 
\add{Existing approaches, such as Groundhog~\cite{Salehnamadi2023groundhog} and BAGEL~\cite{chiou2023bagel}, can be adopted to specifically address the limitation in assessing UI element reachability and interactivity.}
\add{From an interaction trace, LLM-supported} automated checkers can utilize information in the screens immediately before the \add{current} screen to \edit{infer}{acquire} context.
\add{More broadly,} when exploring UI elements on screen, an agent-driven approach can be adopted~\cite{zhang2023appagent, taeb24axnav} to decide whether to further examine an element and how to perform corresponding accessibility action\add{s}.

\subsection{Approximating Real-World Usage}
Although linear navigation is common for screen reader users, people also adopt different ways of interacting with their devices.
As \toolName does not currently support other forms of navigation, accessibility errors are only detected when they present in a linear way.
Touch exploration, for example, does not follow such pattern.
\add{Although crawler-based accessibility evaluation~\cite{eler2018mate, fok_large-scale_chi22} can cover many app screens, they are less efficient and may miss important features in an app, as they typically explore randomly.}
\add{When accessibility evaluation experts audit an app, they carefully select screen samples to make sure they cover an app's ``core purpose and typical user interaction pathways''~\cite{seixas2024exploring}.}
An agent-driven approach can \add{again} be explored to simulate these interactions, \add{learn from how developers and auditors navigate and select screen samples by potentially adopting a cognitive walkthrough approach~\cite{lewis1990testing},} and expose currently undetectable accessibility errors.

\subsection{Direct Code Auditing and Refactoring}
Although linters exist in current development tools~\add{\cite{accesslint, accesslinter, deque2024axelint}}, they are rule-based and limited to the set of accessibility errors they are programmed to detect.
In this paper, we did not address student developer requests for direct code auditing and refactoring.
We intend to explore ways to integrate code-level analysis and support in future iterations of \toolName, \add{such as automatically surfacing code snippets that caused an error, suggesting fixes, and re-evaluating the modified app to confirm repairs}.

\section{Conclusion}
    \toolName presents a novel approach to accessibility assessment in mobile apps through the integration of large language models.
\add{By directly invoking the screen reader and examining its transcripts, \toolName identifies accessibility errors that are previously undetectable with traditional methods, such as grouping and label quality errors, improving coverage from 31\% to 69\%.
\toolName also complements existing rule-based checkers with nuanced evaluations and contextual insights.
Our results uncover opportunities to enhance \toolName by integrating more contextual information, improving interactive context understanding and UI element coverage, and supporting diverse user interactions.
Ultimately, this work sets the stage for aligning automated tools more closely with access technology user needs for inclusive mobile experiences, while opening avenues for integrating LLM-driven heuristic evaluations and simulated user testing to address a wider range of accessibility challenges and user abilities.
}

\begin{acks}
We thank Nigini Oliveira for his generous support in the student developer study,
Douglas Goist for his feedback on system design, and the NSITE team for participant resources.
We thank our participants and anonymous reviewers for their feedback.

This work was supported in part by \grantsponsor{google}{Google}{} and by the \grantsponsor{create}{University of Washington Center for Research and Education on Accessible Technology and Experiences (CREATE)}{}.
Any opinions, findings, conclusions or recommendations expressed in our work are those of the authors and do not necessarily reflect those of any sponsor.
\end{acks}

\bibliographystyle{ACM-Reference-Format}
\bibliography{ref}


\begin{thebibliography}{75}


\ifx \showCODEN    \undefined \def \showCODEN     #1{\unskip}     \fi
\ifx \showISBNx    \undefined \def \showISBNx     #1{\unskip}     \fi
\ifx \showISBNxiii \undefined \def \showISBNxiii  #1{\unskip}     \fi
\ifx \showISSN     \undefined \def \showISSN      #1{\unskip}     \fi
\ifx \showLCCN     \undefined \def \showLCCN      #1{\unskip}     \fi
\ifx \shownote     \undefined \def \shownote      #1{#1}          \fi
\ifx \showarticletitle \undefined \def \showarticletitle #1{#1}   \fi
\ifx \showURL      \undefined \def \showURL       {\relax}        \fi
\providecommand\bibfield[2]{#2}
\providecommand\bibinfo[2]{#2}
\providecommand\natexlab[1]{#1}
\providecommand\showeprint[2][]{arXiv:#2}

\bibitem[Access(2022)]%
        {levelaccess2022report}
\bibfield{author}{\bibinfo{person}{Level Access}.} \bibinfo{year}{2022}\natexlab{}.
\newblock \bibinfo{title}{2022 State of Digital Accessibility}.
\newblock
\urldef\tempurl%
\url{https://www.levelaccess.com/resources/state-of-digital-accessibility-report-2022/}
\showURL{%
\tempurl}


\bibitem[Access(2024)]%
        {levelaccess2024report}
\bibfield{author}{\bibinfo{person}{Level Access}.} \bibinfo{year}{2024}\natexlab{}.
\newblock \bibinfo{title}{Fifth Annual State of Digital Accessibility Report: 2023--2024}.
\newblock
\urldef\tempurl%
\url{https://www.levelaccess.com/resources/fifth-annual-state-of-digital-accessibility-report-2023-2024/}
\showURL{%
\tempurl}


\bibitem[AccessLint(2024)]%
        {accesslint}
\bibfield{author}{\bibinfo{person}{AccessLint}.} \bibinfo{year}{2024}\natexlab{}.
\newblock \bibinfo{title}{Automated and continuous web accessibility testing}.
\newblock
\urldef\tempurl%
\url{https://accesslint.com/}
\showURL{%
\tempurl}


\bibitem[Amalfitano et~al\mbox{.}(2019)]%
        {amalfitano2019combining}
\bibfield{author}{\bibinfo{person}{Domenico Amalfitano}, \bibinfo{person}{Vincenzo Riccio}, \bibinfo{person}{Nicola Amatucci}, \bibinfo{person}{Vincenzo~De Simone}, {and} \bibinfo{person}{Anna~Rita Fasolino}.} \bibinfo{year}{2019}\natexlab{}.
\newblock \showarticletitle{Combining Automated GUI Exploration of Android apps with Capture and Replay through Machine Learning}.
\newblock \bibinfo{journal}{\emph{Information and Software Technology}}  \bibinfo{volume}{105} (\bibinfo{year}{2019}), \bibinfo{pages}{95--116}.
\newblock
\showISSN{0950-5849}
\href{https://doi.org/10.1016/j.infsof.2018.08.007}{doi:\nolinkurl{10.1016/j.infsof.2018.08.007}}


\bibitem[Anthropic(2024)]%
        {claude35sonnet}
\bibfield{author}{\bibinfo{person}{Anthropic}.} \bibinfo{year}{2024}\natexlab{}.
\newblock \bibinfo{title}{Claude 3.5 Sonnet}.
\newblock
\urldef\tempurl%
\url{https://www.anthropic.com/claude/sonnet}
\showURL{%
\tempurl}


\bibitem[Apple(2023)]%
        {apple_use_2023}
\bibfield{author}{\bibinfo{person}{Apple}.} \bibinfo{year}{2023}\natexlab{}.
\newblock \bibinfo{title}{Use {VoiceOver} Recognition on your {iPhone} or {iPad}}.
\newblock
\urldef\tempurl%
\url{https://support.apple.com/en-us/111799}
\showURL{%
\tempurl}


\bibitem[Apple(2024a)]%
        {a11y_inspector}
\bibfield{author}{\bibinfo{person}{Apple}.} \bibinfo{year}{2024}\natexlab{a}.
\newblock \bibinfo{title}{Accessibility Inspector}.
\newblock
\urldef\tempurl%
\url{https://developer.apple.com/documentation/accessibility/accessibility-inspector}
\showURL{%
\tempurl}


\bibitem[Apple(2024b)]%
        {apple_guidelines}
\bibfield{author}{\bibinfo{person}{Apple}.} \bibinfo{year}{2024}\natexlab{b}.
\newblock \bibinfo{title}{Human Interface Guidelines - Accessibility}.
\newblock
\urldef\tempurl%
\url{https://developer.apple.com/design/human-interface-guidelines/accessibility}
\showURL{%
\tempurl}


\bibitem[Braun and Clarke(2006)]%
        {braun2006thematic}
\bibfield{author}{\bibinfo{person}{Virginia Braun} {and} \bibinfo{person}{Victoria Clarke}.} \bibinfo{year}{2006}\natexlab{}.
\newblock \showarticletitle{Using thematic analysis in psychology}.
\newblock \bibinfo{journal}{\emph{Qualitative research in psychology}} \bibinfo{volume}{3}, \bibinfo{number}{2} (\bibinfo{year}{2006}), \bibinfo{pages}{77--101}.
\newblock


\bibitem[Carvalho et~al\mbox{.}(2018)]%
        {Carvalho_2018_a11yproblems}
\bibfield{author}{\bibinfo{person}{Michael Crystian~Nepomuceno Carvalho}, \bibinfo{person}{Felipe~Silva Dias}, \bibinfo{person}{Aline Grazielle~Silva Reis}, {and} \bibinfo{person}{Andr\'{e}~Pimenta Freire}.} \bibinfo{year}{2018}\natexlab{}.
\newblock \showarticletitle{Accessibility and usability problems encountered on websites and applications in mobile devices by blind and normal-vision users}. In \bibinfo{booktitle}{\emph{Proceedings of the 33rd Annual ACM Symposium on Applied Computing}} (Pau, France) \emph{(\bibinfo{series}{SAC '18})}. \bibinfo{publisher}{Association for Computing Machinery}, \bibinfo{address}{New York, NY, USA}, \bibinfo{pages}{2022–2029}.
\newblock
\showISBNx{9781450351911}
\href{https://doi.org/10.1145/3167132.3167349}{doi:\nolinkurl{10.1145/3167132.3167349}}


\bibitem[Casare et~al\mbox{.}(2016)]%
        {casare2016usability}
\bibfield{author}{\bibinfo{person}{Andreia~R. Casare}, \bibinfo{person}{Celmar~G. da Silva}, \bibinfo{person}{Paulo~S. Martins}, {and} \bibinfo{person}{Regina L.~O. Moraes}.} \bibinfo{year}{2016}\natexlab{}.
\newblock \showarticletitle{Usability heuristics and accessibility guidelines: a comparison of heuristic evaluation and WCAG}. In \bibinfo{booktitle}{\emph{Proceedings of the 31st Annual ACM Symposium on Applied Computing}} (Pisa, Italy) \emph{(\bibinfo{series}{SAC '16})}. \bibinfo{publisher}{Association for Computing Machinery}, \bibinfo{address}{New York, NY, USA}, \bibinfo{pages}{213–215}.
\newblock
\showISBNx{9781450337397}
\href{https://doi.org/10.1145/2851613.2851913}{doi:\nolinkurl{10.1145/2851613.2851913}}


\bibitem[Chang et~al\mbox{.}(2024)]%
        {chang2024measuring}
\bibfield{author}{\bibinfo{person}{Trenton Chang}, \bibinfo{person}{Jenna Wiens}, \bibinfo{person}{Tobias Schnabel}, {and} \bibinfo{person}{Adith Swaminathan}.} \bibinfo{year}{2024}\natexlab{}.
\newblock \showarticletitle{Measuring Steerability in Large Language Models}. In \bibinfo{booktitle}{\emph{Neurips Safe Generative AI Workshop 2024}}. \bibinfo{numpages}{19}~pages.
\newblock
\urldef\tempurl%
\url{https://openreview.net/forum?id=y2J5dAqcJW}
\showURL{%
\tempurl}


\bibitem[Chen et~al\mbox{.}(2022)]%
        {chen2022accessible}
\bibfield{author}{\bibinfo{person}{Sen Chen}, \bibinfo{person}{Chunyang Chen}, \bibinfo{person}{Lingling Fan}, \bibinfo{person}{Mingming Fan}, \bibinfo{person}{Xian Zhan}, {and} \bibinfo{person}{Yang Liu}.} \bibinfo{year}{2022}\natexlab{}.
\newblock \showarticletitle{Accessible or Not? An Empirical Investigation of Android App Accessibility}.
\newblock \bibinfo{journal}{\emph{IEEE Transactions on Software Engineering}} \bibinfo{volume}{48}, \bibinfo{number}{10} (\bibinfo{year}{2022}), \bibinfo{pages}{3954--3968}.
\newblock
\href{https://doi.org/10.1109/TSE.2021.3108162}{doi:\nolinkurl{10.1109/TSE.2021.3108162}}


\bibitem[Chiou et~al\mbox{.}(2023)]%
        {chiou2023bagel}
\bibfield{author}{\bibinfo{person}{Paul~T. Chiou}, \bibinfo{person}{Ali~S. Alotaibi}, {and} \bibinfo{person}{William~G.J. Halfond}.} \bibinfo{year}{2023}\natexlab{}.
\newblock \showarticletitle{BAGEL: An Approach to Automatically Detect Navigation-Based Web Accessibility Barriers for Keyboard Users}. In \bibinfo{booktitle}{\emph{Proceedings of the 2023 CHI Conference on Human Factors in Computing Systems}} (Hamburg, Germany) \emph{(\bibinfo{series}{CHI '23})}. \bibinfo{publisher}{Association for Computing Machinery}, \bibinfo{address}{New York, NY, USA}, Article \bibinfo{articleno}{45}, \bibinfo{numpages}{17}~pages.
\newblock
\showISBNx{9781450394215}
\href{https://doi.org/10.1145/3544548.3580749}{doi:\nolinkurl{10.1145/3544548.3580749}}


\bibitem[DeepMind(2024)]%
        {gemini}
\bibfield{author}{\bibinfo{person}{Google DeepMind}.} \bibinfo{year}{2024}\natexlab{}.
\newblock \bibinfo{title}{Gemini Pro}.
\newblock
\urldef\tempurl%
\url{https://deepmind.google/technologies/gemini/pro/}
\showURL{%
\tempurl}


\bibitem[Deque(2021)]%
        {axe}
\bibfield{author}{\bibinfo{person}{Deque}.} \bibinfo{year}{2021}\natexlab{}.
\newblock \bibinfo{title}{axe: Accessibility Testing Tools and Software}.
\newblock
\urldef\tempurl%
\url{https://www.deque.com/axe/}
\showURL{%
\tempurl}


\bibitem[Deque(2024a)]%
        {deque2024coverage}
\bibfield{author}{\bibinfo{person}{Deque}.} \bibinfo{year}{2024}\natexlab{a}.
\newblock \bibinfo{title}{The Automated Accessibility Coverage Report}.
\newblock
\urldef\tempurl%
\url{https://accessibility.deque.com/hubfs/Accessibility-Coverage-Report.pdf}
\showURL{%
\tempurl}


\bibitem[Deque(2024b)]%
        {deque2024axelint}
\bibfield{author}{\bibinfo{person}{Deque}.} \bibinfo{year}{2024}\natexlab{b}.
\newblock \bibinfo{title}{axe DevTools Accessibility Linter}.
\newblock
\urldef\tempurl%
\url{https://www.deque.com/axe/devtools/linter/}
\showURL{%
\tempurl}


\bibitem[Deque(2024c)]%
        {axelinterrules}
\bibfield{author}{\bibinfo{person}{Deque}.} \bibinfo{year}{2024}\natexlab{c}.
\newblock \bibinfo{title}{axe DevTools Linter Accessibility Rules}.
\newblock
\urldef\tempurl%
\url{https://docs.deque.com/linter/4.0.0/en/axe-linter-rules}
\showURL{%
\tempurl}


\bibitem[Duan et~al\mbox{.}(2024)]%
        {duanUICritEnhancingAutomated2024}
\bibfield{author}{\bibinfo{person}{Peitong Duan}, \bibinfo{person}{Chin-Yi Cheng}, \bibinfo{person}{Gang Li}, \bibinfo{person}{Bjoern Hartmann}, {and} \bibinfo{person}{Yang Li}.} \bibinfo{year}{2024}\natexlab{}.
\newblock \showarticletitle{UICrit: Enhancing Automated Design Evaluation with a UI Critique Dataset}. In \bibinfo{booktitle}{\emph{Proceedings of the 37th Annual ACM Symposium on User Interface Software and Technology}} (Pittsburgh, PA, USA) \emph{(\bibinfo{series}{UIST '24})}. \bibinfo{publisher}{Association for Computing Machinery}, \bibinfo{address}{New York, NY, USA}, Article \bibinfo{articleno}{46}, \bibinfo{numpages}{17}~pages.
\newblock
\showISBNx{9798400706288}
\href{https://doi.org/10.1145/3654777.3676381}{doi:\nolinkurl{10.1145/3654777.3676381}}


\bibitem[Dubey et~al\mbox{.}(2024)]%
        {dubey2024llama3herdmodels}
\bibfield{author}{\bibinfo{person}{Abhimanyu Dubey}, \bibinfo{person}{Abhinav Jauhri}, \bibinfo{person}{Abhinav Pandey}, \bibinfo{person}{Abhishek Kadian}, \bibinfo{person}{Ahmad Al-Dahle}, \bibinfo{person}{Aiesha Letman}, \bibinfo{person}{Akhil Mathur}, \bibinfo{person}{Alan Schelten}, \bibinfo{person}{Amy Yang}, \bibinfo{person}{Angela Fan}, {et~al\mbox{.}}} \bibinfo{year}{2024}\natexlab{}.
\newblock \bibinfo{title}{The Llama 3 Herd of Models}.
\newblock
\showeprint[arxiv]{2407.21783}~[cs.AI]
\urldef\tempurl%
\url{https://arxiv.org/abs/2407.21783}
\showURL{%
\tempurl}


\bibitem[Eler et~al\mbox{.}(2018)]%
        {eler2018mate}
\bibfield{author}{\bibinfo{person}{Marcelo~Medeiros Eler}, \bibinfo{person}{Jose~Miguel Rojas}, \bibinfo{person}{Yan Ge}, {and} \bibinfo{person}{Gordon Fraser}.} \bibinfo{year}{2018}\natexlab{}.
\newblock \showarticletitle{{ Automated Accessibility Testing of Mobile Apps }}. In \bibinfo{booktitle}{\emph{2018 IEEE 11th International Conference on Software Testing, Verification and Validation (ICST)}}. \bibinfo{publisher}{IEEE Computer Society}, \bibinfo{address}{Los Alamitos, CA, USA}, \bibinfo{pages}{116--126}.
\newblock
\href{https://doi.org/10.1109/ICST.2018.00021}{doi:\nolinkurl{10.1109/ICST.2018.00021}}


\bibitem[Fok et~al\mbox{.}(2022)]%
        {fok_large-scale_chi22}
\bibfield{author}{\bibinfo{person}{Raymond Fok}, \bibinfo{person}{Mingyuan Zhong}, \bibinfo{person}{Anne~Spencer Ross}, \bibinfo{person}{James Fogarty}, {and} \bibinfo{person}{Jacob~O. Wobbrock}.} \bibinfo{year}{2022}\natexlab{}.
\newblock \showarticletitle{A Large-Scale Longitudinal Analysis of Missing Label Accessibility Failures in Android Apps}. In \bibinfo{booktitle}{\emph{Proceedings of the 2022 CHI Conference on Human Factors in Computing Systems}} (New Orleans, LA, USA) \emph{(\bibinfo{series}{CHI '22})}. \bibinfo{publisher}{Association for Computing Machinery}, \bibinfo{address}{New York, NY, USA}, Article \bibinfo{articleno}{461}, \bibinfo{numpages}{16}~pages.
\newblock
\showISBNx{9781450391573}
\href{https://doi.org/10.1145/3491102.3502143}{doi:\nolinkurl{10.1145/3491102.3502143}}


\bibitem[Foundation(2024)]%
        {appt_guidelines}
\bibfield{author}{\bibinfo{person}{Appt Foundation}.} \bibinfo{year}{2024}\natexlab{}.
\newblock \bibinfo{title}{Beginners' Guide to Accessibility Testing}.
\newblock
\urldef\tempurl%
\url{https://appt.org/en/guidelines/beginnersguide-accessibility-testing}
\showURL{%
\tempurl}


\bibitem[GILMOUR et~al\mbox{.}(1985)]%
        {gilmour1985analysis}
\bibfield{author}{\bibinfo{person}{A.~R. GILMOUR}, \bibinfo{person}{R.~D. ANDERSON}, {and} \bibinfo{person}{A.~L. RAE}.} \bibinfo{year}{1985}\natexlab{}.
\newblock \showarticletitle{The analysis of binomial data by a generalized linear mixed model}.
\newblock \bibinfo{journal}{\emph{Biometrika}} \bibinfo{volume}{72}, \bibinfo{number}{3} (\bibinfo{date}{12} \bibinfo{year}{1985}), \bibinfo{pages}{593--599}.
\newblock
\showISSN{0006-3444}
\href{https://doi.org/10.1093/biomet/72.3.593}{doi:\nolinkurl{10.1093/biomet/72.3.593}}
\showeprint{https://academic.oup.com/biomet/article-pdf/72/3/593/704911/72-3-593.pdf}


\bibitem[Google(2024)]%
        {google-atf}
\bibfield{author}{\bibinfo{person}{Google}.} \bibinfo{year}{2024}\natexlab{}.
\newblock \bibinfo{title}{Accessibility Test Framework for Android}.
\newblock
\urldef\tempurl%
\url{https://github.com/google/Accessibility-Test-Framework-for-Android}
\showURL{%
\tempurl}


\bibitem[{Google}(2024)]%
        {talkback}
\bibfield{author}{\bibinfo{person}{{Google}}.} \bibinfo{year}{2024}\natexlab{}.
\newblock \bibinfo{title}{Get started on Android with Talkback}.
\newblock
\urldef\tempurl%
\url{https://support.google.com/accessibility/android/answer/6283677?hl=en}
\showURL{%
\tempurl}


\bibitem[Google(2024a)]%
        {a11y_scanner}
\bibfield{author}{\bibinfo{person}{Google}.} \bibinfo{year}{2024}\natexlab{a}.
\newblock \bibinfo{title}{Get started with Accessibility Scanner}.
\newblock
\urldef\tempurl%
\url{https://support.google.com/accessibility/android/answer/6376570}
\showURL{%
\tempurl}


\bibitem[Google(2024b)]%
        {google_guidelines}
\bibfield{author}{\bibinfo{person}{Google}.} \bibinfo{year}{2024}\natexlab{b}.
\newblock \bibinfo{title}{Make apps more accessible}.
\newblock
\urldef\tempurl%
\url{https://developer.android.com/guide/topics/ui/accessibility/apps}
\showURL{%
\tempurl}


\bibitem[Google(2024c)]%
        {material_design_guidelines}
\bibfield{author}{\bibinfo{person}{Google}.} \bibinfo{year}{2024}\natexlab{c}.
\newblock \bibinfo{title}{Material Design - Accessibility}.
\newblock
\urldef\tempurl%
\url{https://material.io/design/usability/accessibility.html}
\showURL{%
\tempurl}


\bibitem[Google(2024d)]%
        {google_talkback_source_2024}
\bibfield{author}{\bibinfo{person}{Google}.} \bibinfo{year}{2024}\natexlab{d}.
\newblock \bibinfo{title}{{TalkBack} source code}.
\newblock
\urldef\tempurl%
\url{https://github.com/google/talkback}
\showURL{%
\tempurl}


\bibitem[Google(2024e)]%
        {accesslinter}
\bibfield{author}{\bibinfo{person}{Google}.} \bibinfo{year}{2024}\natexlab{e}.
\newblock \bibinfo{title}{Test your app's accessibility}.
\newblock
\urldef\tempurl%
\url{https://developer.android.com/guide/topics/ui/accessibility/testing}
\showURL{%
\tempurl}


\bibitem[Holm(1979)]%
        {holm_bonferroni}
\bibfield{author}{\bibinfo{person}{Sture Holm}.} \bibinfo{year}{1979}\natexlab{}.
\newblock \showarticletitle{A Simple Sequentially Rejective Multiple Test Procedure}.
\newblock \bibinfo{journal}{\emph{Scandinavian Journal of Statistics}} \bibinfo{volume}{6}, \bibinfo{number}{2} (\bibinfo{year}{1979}), \bibinfo{pages}{65--70}.
\newblock
\showISSN{03036898, 14679469}


\bibitem[Huang et~al\mbox{.}(2024a)]%
        {huangAutomaticMacroMining2024}
\bibfield{author}{\bibinfo{person}{Forrest Huang}, \bibinfo{person}{Gang Li}, \bibinfo{person}{Tao Li}, {and} \bibinfo{person}{Yang Li}.} \bibinfo{year}{2024}\natexlab{a}.
\newblock \showarticletitle{Automatic Macro Mining from Interaction Traces at Scale}. In \bibinfo{booktitle}{\emph{Proceedings of the 2024 CHI Conference on Human Factors in Computing Systems}} (Honolulu, HI, USA) \emph{(\bibinfo{series}{CHI '24})}. \bibinfo{publisher}{Association for Computing Machinery}, \bibinfo{address}{New York, NY, USA}, Article \bibinfo{articleno}{1038}, \bibinfo{numpages}{16}~pages.
\newblock
\showISBNx{9798400703300}
\href{https://doi.org/10.1145/3613904.3642074}{doi:\nolinkurl{10.1145/3613904.3642074}}


\bibitem[Huang et~al\mbox{.}(2024b)]%
        {huang2024promptrpa}
\bibfield{author}{\bibinfo{person}{Tian Huang}, \bibinfo{person}{Chun Yu}, \bibinfo{person}{Weinan Shi}, \bibinfo{person}{Zijian Peng}, \bibinfo{person}{David Yang}, \bibinfo{person}{Weiqi Sun}, {and} \bibinfo{person}{Yuanchun Shi}.} \bibinfo{year}{2024}\natexlab{b}.
\newblock \bibinfo{title}{PromptRPA: Generating Robotic Process Automation on Smartphones from Textual Prompts}.
\newblock
\showeprint[arxiv]{2404.02475}~[cs.HC]
\urldef\tempurl%
\url{https://arxiv.org/abs/2404.02475}
\showURL{%
\tempurl}


\bibitem[Kasuga and Yonetani(2024)]%
        {kasuga2024cxsimulator}
\bibfield{author}{\bibinfo{person}{Akira Kasuga} {and} \bibinfo{person}{Ryo Yonetani}.} \bibinfo{year}{2024}\natexlab{}.
\newblock \showarticletitle{CXSimulator: A User Behavior Simulation using LLM Embeddings for Web-Marketing Campaign Assessment}. In \bibinfo{booktitle}{\emph{Proceedings of the 33rd ACM International Conference on Information and Knowledge Management}} (Boise, ID, USA) \emph{(\bibinfo{series}{CIKM '24})}. \bibinfo{publisher}{Association for Computing Machinery}, \bibinfo{address}{New York, NY, USA}, \bibinfo{pages}{3817–3821}.
\newblock
\showISBNx{9798400704369}
\href{https://doi.org/10.1145/3627673.3679894}{doi:\nolinkurl{10.1145/3627673.3679894}}


\bibitem[Lewis et~al\mbox{.}(1990)]%
        {lewis1990testing}
\bibfield{author}{\bibinfo{person}{Clayton Lewis}, \bibinfo{person}{Peter~G. Polson}, \bibinfo{person}{Cathleen Wharton}, {and} \bibinfo{person}{John Rieman}.} \bibinfo{year}{1990}\natexlab{}.
\newblock \showarticletitle{Testing a walkthrough methodology for theory-based design of walk-up-and-use interfaces}. In \bibinfo{booktitle}{\emph{Proceedings of the SIGCHI Conference on Human Factors in Computing Systems}} (Seattle, Washington, USA) \emph{(\bibinfo{series}{CHI '90})}. \bibinfo{publisher}{Association for Computing Machinery}, \bibinfo{address}{New York, NY, USA}, \bibinfo{pages}{235–242}.
\newblock
\showISBNx{0201509326}
\href{https://doi.org/10.1145/97243.97279}{doi:\nolinkurl{10.1145/97243.97279}}


\bibitem[Liang et~al\mbox{.}(2024)]%
        {liangCanLargeLanguage2023a}
\bibfield{author}{\bibinfo{person}{Weixin Liang}, \bibinfo{person}{Yuhui Zhang}, \bibinfo{person}{Hancheng Cao}, \bibinfo{person}{Binglu Wang}, \bibinfo{person}{Daisy~Yi Ding}, \bibinfo{person}{Xinyu Yang}, \bibinfo{person}{Kailas Vodrahalli}, \bibinfo{person}{Siyu He}, \bibinfo{person}{Daniel~Scott Smith}, \bibinfo{person}{Yian Yin}, \bibinfo{person}{Daniel~A. McFarland}, {and} \bibinfo{person}{James Zou}.} \bibinfo{year}{2024}\natexlab{}.
\newblock \showarticletitle{Can Large Language Models Provide Useful Feedback on Research Papers? A Large-Scale Empirical Analysis}.
\newblock \bibinfo{journal}{\emph{NEJM AI}} \bibinfo{volume}{1}, \bibinfo{number}{8} (\bibinfo{year}{2024}), \bibinfo{pages}{AIoa2400196}.
\newblock
\href{https://doi.org/10.1056/AIoa2400196}{doi:\nolinkurl{10.1056/AIoa2400196}}
\showeprint{https://ai.nejm.org/doi/pdf/10.1056/AIoa2400196}


\bibitem[Liu et~al\mbox{.}(2023)]%
        {liu2023chatting}
\bibfield{author}{\bibinfo{person}{Zhe Liu}, \bibinfo{person}{Chunyang Chen}, \bibinfo{person}{Junjie Wang}, \bibinfo{person}{Mengzhuo Chen}, \bibinfo{person}{Boyu Wu}, \bibinfo{person}{Xing Che}, \bibinfo{person}{Dandan Wang}, {and} \bibinfo{person}{Qing Wang}.} \bibinfo{year}{2023}\natexlab{}.
\newblock \bibinfo{title}{Chatting with GPT-3 for Zero-Shot Human-Like Mobile Automated GUI Testing}.
\newblock
\showeprint[arxiv]{2305.09434}~[cs.SE]
\urldef\tempurl%
\url{https://arxiv.org/abs/2305.09434}
\showURL{%
\tempurl}


\bibitem[MagentaA11y(2024)]%
        {magentaa11y}
\bibfield{author}{\bibinfo{person}{MagentaA11y}.} \bibinfo{year}{2024}\natexlab{}.
\newblock \bibinfo{title}{Native app accessibility checklist}.
\newblock
\urldef\tempurl%
\url{https://www.magentaa11y.com/native/}
\showURL{%
\tempurl}


\bibitem[Mateus et~al\mbox{.}(2020)]%
        {mateus2020eval}
\bibfield{author}{\bibinfo{person}{Delvani~Ant\^{o}nio Mateus}, \bibinfo{person}{Carlos~Alberto Silva}, \bibinfo{person}{Marcelo~Medeiros Eler}, {and} \bibinfo{person}{Andr\'{e}~Pimenta Freire}.} \bibinfo{year}{2020}\natexlab{}.
\newblock \showarticletitle{Accessibility of mobile applications: evaluation by users with visual impairment and by automated tools}. In \bibinfo{booktitle}{\emph{Proceedings of the 19th Brazilian Symposium on Human Factors in Computing Systems}} (Diamantina, Brazil) \emph{(\bibinfo{series}{IHC '20})}. \bibinfo{publisher}{Association for Computing Machinery}, \bibinfo{address}{New York, NY, USA}, Article \bibinfo{articleno}{4}, \bibinfo{numpages}{10}~pages.
\newblock
\showISBNx{9781450381727}
\href{https://doi.org/10.1145/3424953.3426633}{doi:\nolinkurl{10.1145/3424953.3426633}}


\bibitem[Mehralian et~al\mbox{.}(2023)]%
        {Mehralian2023overly}
\bibfield{author}{\bibinfo{person}{Forough Mehralian}, \bibinfo{person}{Navid Salehnamadi}, \bibinfo{person}{Syed~Fatiul Huq}, {and} \bibinfo{person}{Sam Malek}.} \bibinfo{year}{2023}\natexlab{}.
\newblock \showarticletitle{Too Much Accessibility is Harmful! Automated Detection and Analysis of Overly Accessible Elements in Mobile Apps}. In \bibinfo{booktitle}{\emph{Proceedings of the 37th IEEE/ACM International Conference on Automated Software Engineering}} (Rochester, MI, USA) \emph{(\bibinfo{series}{ASE '22})}. \bibinfo{publisher}{Association for Computing Machinery}, \bibinfo{address}{New York, NY, USA}, Article \bibinfo{articleno}{103}, \bibinfo{numpages}{13}~pages.
\newblock
\showISBNx{9781450394758}
\href{https://doi.org/10.1145/3551349.3560424}{doi:\nolinkurl{10.1145/3551349.3560424}}


\bibitem[Nearform(2024)]%
        {reactnative_eslint}
\bibfield{author}{\bibinfo{person}{Nearform}.} \bibinfo{year}{2024}\natexlab{}.
\newblock \bibinfo{title}{Eslint-plugin-react-native-a11y}.
\newblock
\urldef\tempurl%
\url{https://github.com/FormidableLabs/eslint-plugin-react-native-a11y}
\showURL{%
\tempurl}


\bibitem[Nielsen and Molich(1990)]%
        {nielson1990heuristic}
\bibfield{author}{\bibinfo{person}{Jakob Nielsen} {and} \bibinfo{person}{Rolf Molich}.} \bibinfo{year}{1990}\natexlab{}.
\newblock \showarticletitle{Heuristic evaluation of user interfaces}. In \bibinfo{booktitle}{\emph{Proceedings of the SIGCHI Conference on Human Factors in Computing Systems}} (Seattle, Washington, USA) \emph{(\bibinfo{series}{CHI '90})}. \bibinfo{publisher}{Association for Computing Machinery}, \bibinfo{address}{New York, NY, USA}, \bibinfo{pages}{249–256}.
\newblock
\showISBNx{0201509326}
\href{https://doi.org/10.1145/97243.97281}{doi:\nolinkurl{10.1145/97243.97281}}


\bibitem[{OpenAI}(2023)]%
        {openai_gpt-4_2023}
\bibfield{author}{\bibinfo{person}{{OpenAI}}.} \bibinfo{year}{2023}\natexlab{}.
\newblock \bibinfo{title}{{GPT}-4 Technical Report}.
\newblock
\showeprint[arxiv]{2303.08774}
\urldef\tempurl%
\url{http://arxiv.org/abs/2303.08774}
\showURL{%
\tempurl}


\bibitem[OpenAI(2024a)]%
        {gpt4o}
\bibfield{author}{\bibinfo{person}{OpenAI}.} \bibinfo{year}{2024}\natexlab{a}.
\newblock \bibinfo{title}{GPT-4o}.
\newblock
\urldef\tempurl%
\url{https://openai.com/index/hello-gpt-4o/}
\showURL{%
\tempurl}


\bibitem[OpenAI(2024b)]%
        {openai-o1}
\bibfield{author}{\bibinfo{person}{OpenAI}.} \bibinfo{year}{2024}\natexlab{b}.
\newblock \bibinfo{title}{Introducing OpenAI o1}.
\newblock
\urldef\tempurl%
\url{https://openai.com/o1/}
\showURL{%
\tempurl}


\bibitem[Pellegrini et~al\mbox{.}(2020)]%
        {pellegrini2020prioritize}
\bibfield{author}{\bibinfo{person}{Fernanda Pellegrini}, \bibinfo{person}{Marcelo Anjos}, \bibinfo{person}{Fabiana Florentin}, \bibinfo{person}{Bruno Ribeiro}, \bibinfo{person}{Walter Correia}, {and} \bibinfo{person}{Jonysberg Quintino}.} \bibinfo{year}{2020}\natexlab{}.
\newblock \showarticletitle{How to Prioritize Accessibility in Agile Projects}. In \bibinfo{booktitle}{\emph{Advances in Usability and User Experience}}, \bibfield{editor}{\bibinfo{person}{Tareq Ahram} {and} \bibinfo{person}{Christianne Falc{\~a}o}} (Eds.). \bibinfo{publisher}{Springer International Publishing}, \bibinfo{address}{Cham}, \bibinfo{pages}{271--280}.
\newblock
\showISBNx{978-3-030-19135-1}


\bibitem[Power et~al\mbox{.}(2012)]%
        {power2012_guidelines}
\bibfield{author}{\bibinfo{person}{Christopher Power}, \bibinfo{person}{Andr\'{e} Freire}, \bibinfo{person}{Helen Petrie}, {and} \bibinfo{person}{David Swallow}.} \bibinfo{year}{2012}\natexlab{}.
\newblock \showarticletitle{Guidelines are only half of the story: accessibility problems encountered by blind users on the web}. In \bibinfo{booktitle}{\emph{Proceedings of the SIGCHI Conference on Human Factors in Computing Systems}} (Austin, Texas, USA) \emph{(\bibinfo{series}{CHI '12})}. \bibinfo{publisher}{Association for Computing Machinery}, \bibinfo{address}{New York, NY, USA}, \bibinfo{pages}{433–442}.
\newblock
\showISBNx{9781450310154}
\href{https://doi.org/10.1145/2207676.2207736}{doi:\nolinkurl{10.1145/2207676.2207736}}


\bibitem[Recommendation(2021)]%
        {w3c_wcag2.1}
\bibfield{author}{\bibinfo{person}{W3C World Wide Web~Consortium Recommendation}.} \bibinfo{year}{2021}\natexlab{}.
\newblock \bibinfo{title}{Web Content Accessibility Guidelines 2.1}.
\newblock
\urldef\tempurl%
\url{https://www.w3.org/TR/WCAG21/}
\showURL{%
\tempurl}


\bibitem[Rosmaita(2006)]%
        {accessibility_first}
\bibfield{author}{\bibinfo{person}{Brian~J. Rosmaita}.} \bibinfo{year}{2006}\natexlab{}.
\newblock \showarticletitle{Accessibility first! a new approach to web design}. In \bibinfo{booktitle}{\emph{Proceedings of the 37th SIGCSE Technical Symposium on Computer Science Education}} (Houston, Texas, USA) \emph{(\bibinfo{series}{SIGCSE '06})}. \bibinfo{publisher}{Association for Computing Machinery}, \bibinfo{address}{New York, NY, USA}, \bibinfo{pages}{270–274}.
\newblock
\showISBNx{1595932593}
\href{https://doi.org/10.1145/1121341.1121426}{doi:\nolinkurl{10.1145/1121341.1121426}}


\bibitem[Ross et~al\mbox{.}(2018)]%
        {ross_examining_assets18}
\bibfield{author}{\bibinfo{person}{Anne~Spencer Ross}, \bibinfo{person}{Xiaoyi Zhang}, \bibinfo{person}{James Fogarty}, {and} \bibinfo{person}{Jacob~O. Wobbrock}.} \bibinfo{year}{2018}\natexlab{}.
\newblock \showarticletitle{Examining Image-Based Button Labeling for Accessibility in Android Apps through Large-Scale Analysis}. In \bibinfo{booktitle}{\emph{Proceedings of the 20th International ACM SIGACCESS Conference on Computers and Accessibility}} (Galway, Ireland) \emph{(\bibinfo{series}{ASSETS '18})}. \bibinfo{publisher}{Association for Computing Machinery}, \bibinfo{address}{New York, NY, USA}, \bibinfo{pages}{119–130}.
\newblock
\showISBNx{9781450356503}
\href{https://doi.org/10.1145/3234695.3236364}{doi:\nolinkurl{10.1145/3234695.3236364}}


\bibitem[Ross et~al\mbox{.}(2020)]%
        {ross_epidemiology_taccess20}
\bibfield{author}{\bibinfo{person}{Anne~Spencer Ross}, \bibinfo{person}{Xiaoyi Zhang}, \bibinfo{person}{James Fogarty}, {and} \bibinfo{person}{Jacob~O. Wobbrock}.} \bibinfo{year}{2020}\natexlab{}.
\newblock \showarticletitle{An Epidemiology-inspired Large-scale Analysis of Android App Accessibility}.
\newblock \bibinfo{journal}{\emph{ACM Trans. Access. Comput.}} \bibinfo{volume}{13}, \bibinfo{number}{1}, Article \bibinfo{articleno}{4} (\bibinfo{date}{April} \bibinfo{year}{2020}), \bibinfo{numpages}{36}~pages.
\newblock
\showISSN{1936-7228}
\href{https://doi.org/10.1145/3348797}{doi:\nolinkurl{10.1145/3348797}}


\bibitem[Salehnamadi et~al\mbox{.}(2023)]%
        {Salehnamadi2023groundhog}
\bibfield{author}{\bibinfo{person}{Navid Salehnamadi}, \bibinfo{person}{Forough Mehralian}, {and} \bibinfo{person}{Sam Malek}.} \bibinfo{year}{2023}\natexlab{}.
\newblock \showarticletitle{Groundhog: An Automated Accessibility Crawler for Mobile Apps}. In \bibinfo{booktitle}{\emph{Proceedings of the 37th IEEE/ACM International Conference on Automated Software Engineering}} (Rochester, MI, USA) \emph{(\bibinfo{series}{ASE '22})}. \bibinfo{publisher}{Association for Computing Machinery}, \bibinfo{address}{New York, NY, USA}, Article \bibinfo{articleno}{50}, \bibinfo{numpages}{12}~pages.
\newblock
\showISBNx{9781450394758}
\href{https://doi.org/10.1145/3551349.3556905}{doi:\nolinkurl{10.1145/3551349.3556905}}


\bibitem[Santacruz~Abadiano(2020)]%
        {santacruz2020flutter}
\bibfield{author}{\bibinfo{person}{Camilo Santacruz~Abadiano}.} \bibinfo{year}{2020}\natexlab{}.
\newblock \bibinfo{title}{Flutter accessibility suggestion tool}.
\newblock


\bibitem[Seixas~Pereira et~al\mbox{.}(2024)]%
        {seixas2024exploring}
\bibfield{author}{\bibinfo{person}{Let\'{\i}cia Seixas~Pereira}, \bibinfo{person}{Maria Matos}, {and} \bibinfo{person}{Carlos Duarte}.} \bibinfo{year}{2024}\natexlab{}.
\newblock \showarticletitle{Exploring Mobile Device Accessibility: Challenges, Insights, and Recommendations for Evaluation Methodologies}. In \bibinfo{booktitle}{\emph{Proceedings of the 2024 CHI Conference on Human Factors in Computing Systems}} (Honolulu, HI, USA) \emph{(\bibinfo{series}{CHI '24})}. \bibinfo{publisher}{Association for Computing Machinery}, \bibinfo{address}{New York, NY, USA}, Article \bibinfo{articleno}{964}, \bibinfo{numpages}{17}~pages.
\newblock
\showISBNx{9798400703300}
\href{https://doi.org/10.1145/3613904.3642526}{doi:\nolinkurl{10.1145/3613904.3642526}}


\bibitem[Shirogane et~al\mbox{.}(2011)]%
        {shirogane2011method}
\bibfield{author}{\bibinfo{person}{Junko Shirogane}, \bibinfo{person}{Takayuki Kato}, \bibinfo{person}{Yui Hashimoto}, \bibinfo{person}{Kenji Tachibana}, \bibinfo{person}{Hajime Iwata}, {and} \bibinfo{person}{Yoshiaki Fukazawa}.} \bibinfo{year}{2011}\natexlab{}.
\newblock \showarticletitle{Method to Improve Accessibility of Rich Internet Applications}. In \bibinfo{booktitle}{\emph{Information Quality in E-{{Health}}}}, \bibfield{editor}{\bibinfo{person}{Andreas Holzinger} {and} \bibinfo{person}{Klaus-Martin Simonic}} (Eds.). \bibinfo{publisher}{Springer Berlin Heidelberg}, \bibinfo{address}{Berlin, Heidelberg}, \bibinfo{pages}{349--365}.
\newblock
\showISBNx{978-3-642-25364-5}


\bibitem[Silva et~al\mbox{.}(2018)]%
        {silva2018survey}
\bibfield{author}{\bibinfo{person}{Camila Silva}, \bibinfo{person}{Marcelo~Medeiros Eler}, {and} \bibinfo{person}{Gordon Fraser}.} \bibinfo{year}{2018}\natexlab{}.
\newblock \showarticletitle{A survey on the tool support for the automatic evaluation of mobile accessibility}. In \bibinfo{booktitle}{\emph{Proceedings of the 8th International Conference on Software Development and Technologies for Enhancing Accessibility and Fighting Info-Exclusion}} (Thessaloniki, Greece) \emph{(\bibinfo{series}{DSAI '18})}. \bibinfo{publisher}{Association for Computing Machinery}, \bibinfo{address}{New York, NY, USA}, \bibinfo{pages}{286–293}.
\newblock
\showISBNx{9781450364676}
\href{https://doi.org/10.1145/3218585.3218673}{doi:\nolinkurl{10.1145/3218585.3218673}}


\bibitem[Stephanidis et~al\mbox{.}(1998)]%
        {stephanidis1998universal}
\bibfield{author}{\bibinfo{person}{C Stephanidis}, \bibinfo{person}{D Akoumianakis}, \bibinfo{person}{M Sfyrakis}, {and} \bibinfo{person}{A Paramythis}.} \bibinfo{year}{1998}\natexlab{}.
\newblock \showarticletitle{Universal Accessibility in {{HCI}}: {{Process-oriented}} Design Guidelines and Tool Requirements}. In \bibinfo{booktitle}{\emph{Proceedings of the 4th {{ERCIM Workshop}} on {{User Interfaces}} for All}}. \bibinfo{address}{Stockholm, Sweden}, \bibinfo{numpages}{15}~pages.
\newblock


\bibitem[Stiratelli et~al\mbox{.}(1984)]%
        {robert1984random}
\bibfield{author}{\bibinfo{person}{Robert Stiratelli}, \bibinfo{person}{Nan Laird}, {and} \bibinfo{person}{James~H. Ware}.} \bibinfo{year}{1984}\natexlab{}.
\newblock \showarticletitle{Random-Effects Models for Serial Observations with Binary Response}.
\newblock \bibinfo{journal}{\emph{Biometrics}} \bibinfo{volume}{40}, \bibinfo{number}{4} (\bibinfo{year}{1984}), \bibinfo{pages}{961--971}.
\newblock
\showISSN{0006341X, 15410420}
\urldef\tempurl%
\url{http://www.jstor.org/stable/2531147}
\showURL{%
\tempurl}


\bibitem[Swearngin et~al\mbox{.}(2024)]%
        {swearngin2024towards}
\bibfield{author}{\bibinfo{person}{Amanda Swearngin}, \bibinfo{person}{Jason Wu}, \bibinfo{person}{Xiaoyi Zhang}, \bibinfo{person}{Esteban Gomez}, \bibinfo{person}{Jen Coughenour}, \bibinfo{person}{Rachel Stukenborg}, \bibinfo{person}{Bhavya Garg}, \bibinfo{person}{Greg Hughes}, \bibinfo{person}{Adriana Hilliard}, \bibinfo{person}{Jeffrey~P. Bigham}, {and} \bibinfo{person}{Jeffrey Nichols}.} \bibinfo{year}{2024}\natexlab{}.
\newblock \showarticletitle{Towards Automated Accessibility Report Generation for Mobile Apps}.
\newblock \bibinfo{journal}{\emph{ACM Trans. Comput.-Hum. Interact.}} \bibinfo{volume}{31}, \bibinfo{number}{4}, Article \bibinfo{articleno}{54} (\bibinfo{date}{Sept.} \bibinfo{year}{2024}), \bibinfo{numpages}{44}~pages.
\newblock
\showISSN{1073-0516}
\href{https://doi.org/10.1145/3674967}{doi:\nolinkurl{10.1145/3674967}}


\bibitem[Taeb et~al\mbox{.}(2024)]%
        {taeb24axnav}
\bibfield{author}{\bibinfo{person}{Maryam Taeb}, \bibinfo{person}{Amanda Swearngin}, \bibinfo{person}{Eldon Schoop}, \bibinfo{person}{Ruijia Cheng}, \bibinfo{person}{Yue Jiang}, {and} \bibinfo{person}{Jeffrey Nichols}.} \bibinfo{year}{2024}\natexlab{}.
\newblock \showarticletitle{AXNav: Replaying Accessibility Tests from Natural Language}. In \bibinfo{booktitle}{\emph{Proceedings of the CHI Conference on Human Factors in Computing Systems}} (Honolulu, HI, USA) \emph{(\bibinfo{series}{CHI '24})}. \bibinfo{publisher}{Association for Computing Machinery}, \bibinfo{address}{New York, NY, USA}, Article \bibinfo{articleno}{962}, \bibinfo{numpages}{16}~pages.
\newblock
\showISBNx{9798400703300}
\href{https://doi.org/10.1145/3613904.3642777}{doi:\nolinkurl{10.1145/3613904.3642777}}


\bibitem[Vaswani et~al\mbox{.}(2017)]%
        {attentionisallyouneed}
\bibfield{author}{\bibinfo{person}{Ashish Vaswani}, \bibinfo{person}{Noam Shazeer}, \bibinfo{person}{Niki Parmar}, \bibinfo{person}{Jakob Uszkoreit}, \bibinfo{person}{Llion Jones}, \bibinfo{person}{Aidan~N Gomez}, \bibinfo{person}{{\L}ukasz Kaiser}, {and} \bibinfo{person}{Illia Polosukhin}.} \bibinfo{year}{2017}\natexlab{}.
\newblock \showarticletitle{Attention Is All You Need}. In \bibinfo{booktitle}{\emph{Advances in Neural Information Processing Systems}} (Long Beach, California, USA), \bibfield{editor}{\bibinfo{person}{I.~Guyon}, \bibinfo{person}{U.~Von Luxburg}, \bibinfo{person}{S.~Bengio}, \bibinfo{person}{H.~Wallach}, \bibinfo{person}{R.~Fergus}, \bibinfo{person}{S.~Vishwanathan}, {and} \bibinfo{person}{R.~Garnett}} (Eds.), Vol.~\bibinfo{volume}{30}. \bibinfo{publisher}{Curran Associates, Inc.}, \bibinfo{address}{Red Hook, NY, USA}, \bibinfo{numpages}{11}~pages.
\newblock


\bibitem[W3C(2021)]%
        {w3c_guidelines}
\bibfield{author}{\bibinfo{person}{W3C}.} \bibinfo{year}{2021}\natexlab{}.
\newblock \bibinfo{title}{Mobile Accessibility: How WCAG 2.0 and Other W3C/WAI Guidelines Apply to Mobile}.
\newblock
\urldef\tempurl%
\url{https://www.w3.org/TR/mobile-accessibility-mapping/}
\showURL{%
\tempurl}


\bibitem[Wang et~al\mbox{.}(2024)]%
        {wang2024mobile}
\bibfield{author}{\bibinfo{person}{Junyang Wang}, \bibinfo{person}{Haiyang Xu}, \bibinfo{person}{Jiabo Ye}, \bibinfo{person}{Ming Yan}, \bibinfo{person}{Weizhou Shen}, \bibinfo{person}{Ji Zhang}, \bibinfo{person}{Fei Huang}, {and} \bibinfo{person}{Jitao Sang}.} \bibinfo{year}{2024}\natexlab{}.
\newblock \bibinfo{title}{Mobile-Agent: Autonomous Multi-Modal Mobile Device Agent with Visual Perception}.
\newblock
\showeprint[arxiv]{2401.16158}~[cs.CL]
\urldef\tempurl%
\url{https://arxiv.org/abs/2401.16158}
\showURL{%
\tempurl}


\bibitem[WebAIM(2024)]%
        {wave}
\bibfield{author}{\bibinfo{person}{WebAIM}.} \bibinfo{year}{2024}\natexlab{}.
\newblock \bibinfo{title}{WAVE Web Accessibility Evaluation Tools}.
\newblock
\urldef\tempurl%
\url{https://wave.webaim.org/}
\showURL{%
\tempurl}


\bibitem[Wei et~al\mbox{.}(2023)]%
        {wei2023chainofthought}
\bibfield{author}{\bibinfo{person}{Jason Wei}, \bibinfo{person}{Xuezhi Wang}, \bibinfo{person}{Dale Schuurmans}, \bibinfo{person}{Maarten Bosma}, \bibinfo{person}{Brian Ichter}, \bibinfo{person}{Fei Xia}, \bibinfo{person}{Ed Chi}, \bibinfo{person}{Quoc Le}, {and} \bibinfo{person}{Denny Zhou}.} \bibinfo{year}{2023}\natexlab{}.
\newblock \bibinfo{title}{Chain-of-Thought Prompting Elicits Reasoning in Large Language Models}.
\newblock
\showeprint[arxiv]{2201.11903}~[cs.CL]
\urldef\tempurl%
\url{https://arxiv.org/abs/2201.11903}
\showURL{%
\tempurl}


\bibitem[Wen et~al\mbox{.}(2024)]%
        {wen2024autodroid}
\bibfield{author}{\bibinfo{person}{Hao Wen}, \bibinfo{person}{Yuanchun Li}, \bibinfo{person}{Guohong Liu}, \bibinfo{person}{Shanhui Zhao}, \bibinfo{person}{Tao Yu}, \bibinfo{person}{Toby Jia-Jun Li}, \bibinfo{person}{Shiqi Jiang}, \bibinfo{person}{Yunhao Liu}, \bibinfo{person}{Yaqin Zhang}, {and} \bibinfo{person}{Yunxin Liu}.} \bibinfo{year}{2024}\natexlab{}.
\newblock \showarticletitle{AutoDroid: LLM-powered Task Automation in Android}. In \bibinfo{booktitle}{\emph{Proceedings of the 30th Annual International Conference on Mobile Computing and Networking}} (Washington D.C., DC, USA) \emph{(\bibinfo{series}{ACM MobiCom '24})}. \bibinfo{publisher}{Association for Computing Machinery}, \bibinfo{address}{New York, NY, USA}, \bibinfo{pages}{543–557}.
\newblock
\showISBNx{9798400704895}
\href{https://doi.org/10.1145/3636534.3649379}{doi:\nolinkurl{10.1145/3636534.3649379}}


\bibitem[Wu et~al\mbox{.}(2024)]%
        {wu2024uiclip}
\bibfield{author}{\bibinfo{person}{Jason Wu}, \bibinfo{person}{Yi-Hao Peng}, \bibinfo{person}{Xin Yue~Amanda Li}, \bibinfo{person}{Amanda Swearngin}, \bibinfo{person}{Jeffrey~P Bigham}, {and} \bibinfo{person}{Jeffrey Nichols}.} \bibinfo{year}{2024}\natexlab{}.
\newblock \showarticletitle{UIClip: A Data-driven Model for Assessing User Interface Design}. In \bibinfo{booktitle}{\emph{Proceedings of the 37th Annual ACM Symposium on User Interface Software and Technology}} (Pittsburgh, PA, USA) \emph{(\bibinfo{series}{UIST '24})}. \bibinfo{publisher}{Association for Computing Machinery}, \bibinfo{address}{New York, NY, USA}, Article \bibinfo{articleno}{45}, \bibinfo{numpages}{16}~pages.
\newblock
\showISBNx{9798400706288}
\href{https://doi.org/10.1145/3654777.3676408}{doi:\nolinkurl{10.1145/3654777.3676408}}


\bibitem[Wu et~al\mbox{.}(2021)]%
        {wu2021screen}
\bibfield{author}{\bibinfo{person}{Jason Wu}, \bibinfo{person}{Xiaoyi Zhang}, \bibinfo{person}{Jeff Nichols}, {and} \bibinfo{person}{Jeffrey~P Bigham}.} \bibinfo{year}{2021}\natexlab{}.
\newblock \showarticletitle{Screen Parsing: Towards Reverse Engineering of UI Models from Screenshots}. In \bibinfo{booktitle}{\emph{The 34th Annual ACM Symposium on User Interface Software and Technology}} (Virtual Event, USA) \emph{(\bibinfo{series}{UIST '21})}. \bibinfo{publisher}{Association for Computing Machinery}, \bibinfo{address}{New York, NY, USA}, \bibinfo{pages}{470–483}.
\newblock
\showISBNx{9781450386357}
\href{https://doi.org/10.1145/3472749.3474763}{doi:\nolinkurl{10.1145/3472749.3474763}}


\bibitem[Xiang et~al\mbox{.}(2024)]%
        {xiang2024simuser}
\bibfield{author}{\bibinfo{person}{Wei Xiang}, \bibinfo{person}{Hanfei Zhu}, \bibinfo{person}{Suqi Lou}, \bibinfo{person}{Xinli Chen}, \bibinfo{person}{Zhenghua Pan}, \bibinfo{person}{Yuping Jin}, \bibinfo{person}{Shi Chen}, {and} \bibinfo{person}{Lingyun Sun}.} \bibinfo{year}{2024}\natexlab{}.
\newblock \showarticletitle{SimUser: Generating Usability Feedback by Simulating Various Users Interacting with Mobile Applications}. In \bibinfo{booktitle}{\emph{Proceedings of the 2024 CHI Conference on Human Factors in Computing Systems}} (Honolulu, HI, USA) \emph{(\bibinfo{series}{CHI '24})}. \bibinfo{publisher}{Association for Computing Machinery}, \bibinfo{address}{New York, NY, USA}, Article \bibinfo{articleno}{9}, \bibinfo{numpages}{17}~pages.
\newblock
\showISBNx{9798400703300}
\href{https://doi.org/10.1145/3613904.3642481}{doi:\nolinkurl{10.1145/3613904.3642481}}


\bibitem[Yan and Ramachandran(2019)]%
        {yan_current_state_accessibility}
\bibfield{author}{\bibinfo{person}{Shunguo Yan} {and} \bibinfo{person}{P.~G. Ramachandran}.} \bibinfo{year}{2019}\natexlab{}.
\newblock \showarticletitle{The Current Status of Accessibility in Mobile Apps}.
\newblock \bibinfo{journal}{\emph{ACM Trans. Access. Comput.}} \bibinfo{volume}{12}, \bibinfo{number}{1}, Article \bibinfo{articleno}{3} (\bibinfo{date}{feb} \bibinfo{year}{2019}), \bibinfo{numpages}{31}~pages.
\newblock
\showISSN{1936-7228}
\href{https://doi.org/10.1145/3300176}{doi:\nolinkurl{10.1145/3300176}}


\bibitem[Zhang et~al\mbox{.}(2023)]%
        {zhang2023appagent}
\bibfield{author}{\bibinfo{person}{Chi Zhang}, \bibinfo{person}{Zhao Yang}, \bibinfo{person}{Jiaxuan Liu}, \bibinfo{person}{Yucheng Han}, \bibinfo{person}{Xin Chen}, \bibinfo{person}{Zebiao Huang}, \bibinfo{person}{Bin Fu}, {and} \bibinfo{person}{Gang Yu}.} \bibinfo{year}{2023}\natexlab{}.
\newblock \bibinfo{title}{AppAgent: Multimodal Agents as Smartphone Users}.
\newblock
\showeprint[arxiv]{2312.13771}~[cs.CV]
\urldef\tempurl%
\url{https://arxiv.org/abs/2312.13771}
\showURL{%
\tempurl}


\bibitem[Zhang et~al\mbox{.}(2024)]%
        {zhangRepairingBugsPython2022}
\bibfield{author}{\bibinfo{person}{Jialu Zhang}, \bibinfo{person}{Jos\'{e}~Pablo Cambronero}, \bibinfo{person}{Sumit Gulwani}, \bibinfo{person}{Vu Le}, \bibinfo{person}{Ruzica Piskac}, \bibinfo{person}{Gustavo Soares}, {and} \bibinfo{person}{Gust Verbruggen}.} \bibinfo{year}{2024}\natexlab{}.
\newblock \showarticletitle{PyDex: Repairing Bugs in Introductory Python Assignments using LLMs}.
\newblock \bibinfo{journal}{\emph{Proc. ACM Program. Lang.}} \bibinfo{volume}{8}, \bibinfo{number}{OOPSLA1}, Article \bibinfo{articleno}{133} (\bibinfo{date}{April} \bibinfo{year}{2024}), \bibinfo{numpages}{25}~pages.
\newblock
\href{https://doi.org/10.1145/3649850}{doi:\nolinkurl{10.1145/3649850}}


\bibitem[Zhong et~al\mbox{.}(2021)]%
        {zhong_helpviz_chi21}
\bibfield{author}{\bibinfo{person}{Mingyuan Zhong}, \bibinfo{person}{Gang Li}, \bibinfo{person}{Peggy Chi}, {and} \bibinfo{person}{Yang Li}.} \bibinfo{year}{2021}\natexlab{}.
\newblock \showarticletitle{HelpViz: Automatic Generation of Contextual Visual Mobile Tutorials from Text-Based Instructions}. In \bibinfo{booktitle}{\emph{The 34th Annual ACM Symposium on User Interface Software and Technology}} (Virtual Event, USA) \emph{(\bibinfo{series}{UIST '21})}. \bibinfo{publisher}{Association for Computing Machinery}, \bibinfo{address}{New York, NY, USA}, \bibinfo{pages}{1144–1153}.
\newblock
\showISBNx{9781450386357}
\href{https://doi.org/10.1145/3472749.3474812}{doi:\nolinkurl{10.1145/3472749.3474812}}


\end{thebibliography}

\appendix

\section{Assignment on Accessibility and Questions Asked in the Assignment}
    \label{sec:appendix-assignment}
    \subsection{Assignment Description}
We provide the full text of the assignment requirement on Accessibility below.
Students had one week to work on the turn in this assignment.

\medskip
\leftskip0.5cm\relax
\rightskip0.5cm\relax
    \noindent
    This week's assignment aims to practice the techniques we discussed regarding Accessibility!

    \bigskip
    \noindent
    \textbf{Step 1}: Choose one of the 3 recent assignments [\textit{assignment numbers}] from this course and examine the accessibility for the app you created.
    
    1.1. Use the Google Accessibility Scanner and notes and screenshots of potential issues.
    
    1.2. Try to access a feature of your App using TalkBack and take notes about difficulties and possible improvements. 
    
    We recommend using a physical device. If you only have access to the emulator, you need to download the Android Accessibility Suite (which requires an emulator with Google Play Store installed).

    \bigskip
    \noindent
    \textbf{Step 2}: Improve the accessibility of your App: refer back to all the tips [\textit{the instructor}] presented in class, and submit a new version OF YOUR CODE with improvements in at least a couple of categories of accessibility mistakes (find them both at the above codelab or the lecture slides.)
    
    By YOUR CODE, we mean you should start this assignment by cloning a previous repository with your own code.

    \bigskip
    \noindent
    \textbf{Submission}:
    You will need to make two separate submissions: (TIP: Make sure you familiarize yourself with the rubric in the reflection part before moving forward with the coding part!)
    
    1. Code: Access your Repo at [\textit{link redacted}]. Submit your code.
    
    2. Reflection: Follow the questions in [\textit{assignment link redacted; see questions below}].

\leftskip0cm\relax
\rightskip0cm\relax

\subsection{Reflection Questions}
Here we provide the full text for the reflection questions asked in the assignment.

\medskip
\leftskip0.5cm\relax
\rightskip0.5cm\relax

    \noindent
    \textbf{I. Issues discovered through the Google Accessibility Scanner}
    
    [\textit{Excluded from our study}] Provide screenshots of the Scanner's results on all of your app's screens, merged together as a single image or PDF.

    \textbf{[Q1]} How helpful was the Accessibility Scanner? Were there unexpected accessibility issues that you discovered by running the scanner? Any ideas on how the Scanner could be improved?

    \bigskip
    \noindent
    \textbf{II. Accessibility issues discovered through TalkBack}
    
    \textbf{[Q2]} Recall the most common types of accessibility errors from [the previous lecture on Accessibility]. Apart from the issues discovered by the Accessibility Scanner, what issues have you discovered using TalkBack?

    \textbf{[Q3]} Reflecting on your experience using TalkBack to interact with your app, what would you do to improve the user experience?

    \bigskip
    \noindent
    \textbf{III. Towards better app accessibility}
    
    \textbf{[Q4]} Reflect on the limitations of the Accessibility Scanner and your experience using TalkBack to test accessibility. We do not want accessibility to be an afterthought. What have Android Studio and Compose done to promote/enforce accessibility in developing apps?

    \textbf{[Q5]} To promote better app accessibility, what else the development tools could do? List a few aspects that would have helped you make your app more accessible in the first place.

\leftskip0cm\relax
\rightskip0cm\relax

\section{Qualitative Analysis of Student Developer Responses}
    \label{sec:appendix-qualitative}
    \subsection{Procedure}
We analyzed responses that students provided to reflection questions in an assignment on Accessibility, available in the previous Appendix section.
This assignment was designed and administered as part of the regular course curriculum.
Specifically, students were instructed to use the Android Accessibility Scanner to check for accessibility issues first, then use TalkBack to confirm them and look for other accessibility problems.
Students were asked to reflect on their experience about analyzing and repairing accessibility issues using the Accessibility Scanner and TalkBack and discuss what else is needed in current developer tools.
These reflections consisted of five questions.

Prior to this assignment, students received seven weeks of instruction on Android development with Jetpack Compose\footnote{Jetpack Compose: \url{https://developer.android.com/compose}} and created six small-scale apps focusing on different aspects of interaction as assignments.
The lectures included two lectures on Android Accessibility's common issues and best practices.
Students were asked to select one of three apps from previous assignments: (1) a Gallery/Image Showcase app, (2) a Dessert Ordering app, (3) a Voice-to-Image Generation app.

This review was conducted as an independent analysis of student response data stripped of all personally identifiable information.
We received formal determination that this project was not human subjects research from the University's IRB.
All data were from undergraduate students, however no other demographic information was available.
In total, 41 anonymized submissions to the assignment were collected.
We excluded submissions where the student indicated that they did not complete all assignment requirements or submitted blank responses, resulting in 36 unique submissions in our analysis.

\subsection{Analysis}
We analyze the following research questions. RQ0 is analyzed to ensure the consistency and validity of student response data when compared to previous research results.
\add{Results for RQ1 and RQ2 are included in Section~\ref{sec:dev-survey}.}

\textbf{RQ0}: What are the types of accessibility failures that the Accessibility Scanner \textit{cannot} identify, but can be identified with TalkBack?

\textbf{RQ1}: What are the problems students had while using current developer tools (including Android Studio, Jetpack Compose, Accessibility Scanner, and TalkBack) to check for accessibility failures?

\textbf{RQ2}: What are the features students wanted to see in the developer tools for improving accessibility?

We used standard thematic analysis processes~\cite{braun2006thematic} to analyze student reflections that corresponded to each of the three research questions.

\subsection{RQ0: Accessibility failures identified by TalkBack only}
We developed eight initial codes which revealed three themes:
(1)~\textit{improper labels and content descriptions},
(2)~\textit{inappropriate feedback for functionalities},
(3)~\textit{unclear structure or navigation order},
(4)~\textit{unexpected TalkBack behavior}.
These issues align with those commonly experienced by blind users but undetectable by automated tools in prior work~\cite{mateus2020eval}.
Specifically, students covered 11 of the 15 frequently encountered problems by blind users.

\subsubsection{Improper labels and content descriptions}
Our first theme shows that accessibility failures in labels and content descriptions are often ignored by the Accessibility Scanner, identifiable only through screen reader testing. 
For example, P5 noted missing labels for buttons, P15 noted improperly labeled buttons, P17 noted improper image alt texts, and P21 noted redundant text in icons:
    \begin{lstlisting}
        For the two buttons "Buy" and "Don't Buy", the announcements were not helpful, as they just said, "Double tap to activate." It doesn't provide any information on what clicking that element will do. However, in my app, understanding what each button does is very crucial. (P5)
    \end{lstlisting}
    \begin{lstlisting}
        The forward and back buttons use a less than, greater than text to signify scrolling through the carousel. However, "less than" and "greater than" buttons, when announced, might not make sense to the user. (P15)
    \end{lstlisting}
    \begin{lstlisting}
        images were not being described with proper alt text, instead, the screenreader can only say "xxx image". (P17)
    \end{lstlisting}
    \begin{lstlisting}
        I had provided description for icons that ended up being repetitive with the subsequent text. (P21)
    \end{lstlisting}

These accessibility failures roughly align with the following frequently encountered problems in~\cite{mateus2020eval}:
(1)~\textit{Lack of identification (Buttons)},
(2)~\textit{Button functionality is unclear or confusing (Buttons)},
(3)~\textit{No textual alternative (Image)}.

\subsubsection{Inappropriate feedback for functionalities}
Our second theme shows that actions that are unannounced or unclear and inconsistent are only discovered after using TalkBack.
For example, P2 found the functionality of the app confusing, P10 found the default label to be inadequate, P31 identified interactions not exposed to screen readers, and P35 noted unannounced states:
    
    \begin{lstlisting}
        The description of the app and images, as well as the buttons are very insufficient, I can't barely understand what this app does. (P2)
    \end{lstlisting}
    \begin{lstlisting}
        I discovered the following issues: 1. share - "Double tap to activate" default click label. (P10)
    \end{lstlisting}
    \begin{lstlisting}
        I have 2 invisible UI interactions: double-clicking the image to like and long pressing to zoom in. These are two interactions that I can't do with TalkBack. Also, their results are invisible to TalkBack or users who would use it. (P31)
    \end{lstlisting}
    \begin{lstlisting}
        The app would not announce the new player upon switching screens, but instead would wait until the user would touch the text which displayed their name. This would be a huge accessibility issue. (P35)
    \end{lstlisting}

These accessibility failures roughly align with the following frequently encountered problems in~\cite{mateus2020eval}:
(1)~\textit{Inappropriate feedback (Controls, forms and functionality)},
(2)~\textit{Unclear or confusing functionality (Controls, forms and functionality)},
(3)~\textit{Default presentation of control or form element is not adequate (Controls, forms and functionality)},
(4)~\textit{Users cannot make sense of content (Content - meaning)}

\subsubsection{Unclear structure or navigation order}
Our third theme shows that inconsistent or confusing navigational orders are only discovered after using screen readers.
P3 noted a lack of navigational feature in their app,
P5 described how overly long interactions are frustrating, and
P11 noted inappropriate navigation sequence.
    \begin{lstlisting}
         Issues I have discovers using TalkBack was that without the shortcut, I didn't know how to leave the page with a swipe up function since there wasn't a button to click and also the swiping up and down didn't work with the TalkBack on. (P3)
    \end{lstlisting}
    \begin{lstlisting}
        The next issue was that in the bottom part of the app where all the statistics are listed (number of desserts sold, dessert price, etc), there were too many focusable elements, which resulted in too much user interaction. It does not make sense for the user to have to interact once to get a description of the statistic (ex: Number of desserts sold) and then interact again to get the actual value (ex: 10). It would be frustrating and tedious to get through all the statistics for each dessert that is shown, especially since knowing the statistics is helpful for the user to make their decision of whether or not to buy the dessert. (P5)
    \end{lstlisting}
    \begin{lstlisting}
        The navigation skipped the next dessert icon, transitioning from "Next Dessert" to "Items Left". (P11)
    \end{lstlisting}
    
These accessibility failures roughly align with the following frequently encountered problems in~\cite{mateus2020eval}:
(1)~\textit{Sequence of interaction is unclear or confusing (Controls, forms and functionality)},
(2)~\textit{Expected functionality not present (Controls, forms and functionality)},
(3)~\textit{Functionality does not work as expected (Controls, forms and functionality)}.

\subsubsection{Unexpected TalkBack behavior}
Our fourth theme shows that unexpected (buggy) TalkBack behaviors were encountered. This aligns with the frequently encountered problem ``\textit{System problems with assistive technology (System characteristic)}''.

Our analysis covered 11 of the 15 frequently encountered problems by blind users. The following four problems are not covered:
(1)~\textit{Users inferred that there was functionality where there wasn’t (Controls, forms and functionality)},
(2)~\textit{Inconsistent Content organization (Content - meaning)},
(3)~\textit{Meaning in content is lost (Content - meaning)},
(4)~\textit{No textual alternative (Audio, video and multimedia)}.

\section{Prompt for \toolName}
    \label{sec:appendix-prompt}
    We provide the prompt \textit{General\_Contextual} used in \toolName.

\begin{lstlisting}
    You are examining the accessibility of an app based on transcripts of TalkBack. You will see a transcript of what a screen reader user will hear when they use an app. Please analyze if each interaction reflected in this transcript is accessible.
    The transcript represents a sequence of TalkBack interactions. Each transcript entry is feedback for one single action.
    
    ## Basics of Accessibility
    We want descriptions to be informative and concise. Consider if the spoken feedback for each element convey its content or purpose appropriately. Most important information should appear first.
    Each description must be unique. That way, when screen reader users encounter a repeated element description, they correctly recognize that the focus is on an element that already had focus earlier. In particular, each item within a view group such as RecyclerView must have a different description. Each description must reflect the content that's unique to a given item, such as the name of a city in a list of locations.
    If the app displays several UI elements that form a natural group, such as details of a song or attributes of a message, these elements should be arranged within a container. This way, accessibility services can present the inner elements' content descriptions, one after the other, in a single announcement. This consolidation of related elements helps users of assistive technology discover the information on the screen more efficiently. Because accessibility services announce the inner elements' descriptions in a single utterance, it's important to keep each description as short as possible while still conveying the element's meaning.
    If an element in the UI exists only for visual spacing or visual appearance purposes, it should not be announced.
    If both type and usage hint are not present, the element is static text. Users will not confuse it with actionable item. This is expected behavior and not an accessibility issue.
    Sometimes a screen contains many elements, and it makes sense to put certain actionable elements at the beginning of TalkBack focus order to make them more discoverable, such as "compose new email" button on a page of inbox messages, or a "close" or "return" button.
    
    ## How to interpret the transcript
    - For each transcript entry, think what the user is doing and what the user is experiencing. For example, if the user is swiping right, the user is moving to the next element. If the user is swiping left, the user is moving to the previous element. If the user is double-tapping, the user is activating the element. 
    - Consider the quality of labels, whether buttons are labeled effectively, whether you could infer what functionalities the interactive elements provide, and overall user experience.
    - Consider the transcript without punctuations, as they are not read aloud. Do not ignore text after the punctuations.
    - Consider each entry in the context of the entries before and after it, as the functionalities and meanings can be associated.
    - Try your best to infer the relationship between entries, and understand the intent of the design. If the inferred intent differs from the presentation in the transcript or that it can be improved, then this may indicate accessibility issue. However, only indicate accessibility issues when they are confusing and impede understanding. Otherwise, indicate as suggestions. When making suggestions, avoid adding text if it can be understood through context.
    - Cosmetic issues, such as capitalization, should not be reported.
    
    ## Your task
    Follow these steps to evaluate the accessibility of the transcript:
    Step 1 - Look at the big picture. Consider each transcript entry in relation to the elements before and after it. Is this element related to nearby elements? If so, do you see accessibility issues that are related to the presentation order and grouping of elements? Should adjacent elements be grouped together? Many accessibility issues can be resolved if items are properly grouped. Enclose all your output for this step within triple quotes (""").
    <@\balance@>
    Step 2 - Based on all guidelines above, consider each transcript entry for its accessibility. What is the intent of each transcript entry? Is the indended meaning converyed through the transcript? Build your analysis on top of Step 1. If an entry has multiple issues, separate them clearly. If an item is uninformative, consider if it needs to be associated with items. Include possible causes when appropriate. Enclose all your output for this step within triple quotes (""").
    Step 3 - Convert your previous analyses into the JSON format specified. Pay attention to index number; be concise; do not remove information. If a same issue is repeated across multiple items, include the issue for EACH occurrence.
    ```
    { 
        "audit": [
            {
                "index": int,
                "transcript": string,   # Copy the full transcript here
                "issue": string,        # Describe the issue here if you have feedback, otherwise leave blank
                "explanation": string,  # Explain the issue in detail here
                "suggestion": string,   # Suggested solution here if mentioned, otherwise leave blank
            },
            ...
        ] 
    }
    ```
\end{lstlisting}

\end{document}